\documentclass[11pt,a4paper]{article}
\usepackage{jheppub}
\usepackage{subfigure}
\usepackage{cancel}
\usepackage{here}
\usepackage{comment,braket}
\usepackage{amsmath}
\usepackage{graphics}
\usepackage{amsfonts}
\usepackage{amssymb}
\usepackage{latexsym}
\usepackage{graphicx}

\preprint{CTPU-PTC-22-04}

\title{Type II Seesaw Leptogenesis}
\author[a]{Neil D. Barrie,}
\author[b]{Chengcheng Han}
\author[c,d,e,f]{and Hitoshi Murayama}
\affiliation[a]{Center for Theoretical Physics of the Universe, Institute for Basic Science (IBS),\\
	Daejeon, 34126, Korea.}
\affiliation[b]{School of Physics, Sun Yat-Sen University,\\
	Guangzhou 510275, China.}
\affiliation[c]{Department of Physics, University of California,\\
	Berkeley, CA 94720, USA.}
\affiliation[d]{Kavli Institute for the Physics and Mathematics of the Universe (WPI), University of Tokyo,\\
	Kashiwa 277-8583, Japan.}
\affiliation[e]{Ernest Orlando Lawrence Berkeley National Laboratory,\\
	Berkeley, CA 94720, USA.}
\affiliation[f]{Hamamatsu Professor}

\emailAdd{nlbarrie@ibs.re.kr}
\emailAdd{hanchch@mail.sysu.edu.cn}
\emailAdd{hitoshi@berkeley.edu, hitoshi.murayama@ipmu.jp}
\abstract{
The Type II Seesaw Mechanism provides a minimal framework to explain the neutrino masses involving the introduction of a single triplet Higgs to the Standard Model. However, this simple extension was believed to be unable to successfully explain the observed baryon asymmetry of the universe through Leptogenesis. In our previous work ({\it Phys.~Rev.~Lett.}~{\bf 128},~141801), we demonstrated that the triplet Higgs of the Type II Seesaw Mechanism alone can simultaneously generate the observed baryon asymmetry of the universe and the neutrino masses while playing a role in setting up Inflation. This is achievable with a triplet Higgs mass as low as 1 TeV, and predicts that the neutral component obtains a small vacuum expectation value $v_\Delta < 10 $ keV. We find that our model has very rich phenomenology and can be tested by various terrestrial experiments as well as by astronomical observations. Particularly, we show that the successful parameter region may be probed at a future 100 TeV collider, upcoming lepton flavor violation experiments such as Mu3e, and neutrinoless double beta decay experiments. Additionally, the tensor-to-scalar ratio from the inflationary scenario will be probed by the LiteBIRD telescope, and observable isocurvature perturbations may be produced for some parameter choices. In this article, we present all the technical details of our calculations and further discussion of its phenomenological implications. }
\arxivnumber{2204.XXXXX}

\begin{document}
    
    \maketitle

\section{Introduction}
One of the most significant mysteries in modern physics is the existence of non-zero neutrino masses. In the Standard Model (SM) the neutrinos are massless by definition, meaning that  extensions to the SM must be proposed to explain their observed mass. The leading mechanism for generating the neutrino masses is known as the Seesaw Mechanism. The Seesaw Mechanism can be split into three main categories called Type I, II and III. The Type I  Seesaw Mechanism is the most widely studied, consisting of an extension of the SM by at least two right-handed neutrinos \cite{Minkowski:1977sc, Yanagida:1979as, Glashow:1979nm, GellMann:1980vs}. The Type II Seesaw Mechanism involves the inclusion of a triplet Higgs to the SM, which produces a Majorana mass term for the neutrinos when its neutral component takes its non-zero vacuum expectation value (VEV) \cite{Magg:1980ut,Cheng:1980qt,Lazarides:1980nt,Mohapatra:1980yp}. This new $ SU(2) $ triplet scalar can have rich terrestrial phenomenology dependent on its mass scale. On the other hand, the Type III Seesaw Mechanism requires the existence of at least two $ SU(2) $ fermion triplets \cite{Foot:1988aq,Albright:2003xb}. None of these models, or mixtures of them, have been experimentally observed and theoretical work regarding their phenomenological implications is ongoing. Planned future experiments will probe the associated particles and bring us closer to understanding the origin of the neutrino masses. 

An important feature of the neutrino sector is that it may be closely related to another unexplained observation: the baryon asymmetry of our universe. Thanks to the sphaleron process~\cite{Kuzmin:1985mm}, if a lepton asymmetry can be generated before the Electroweak Phase Transition (EWPT), it will be transferred to the baryonic sector via the sphalerons~\cite{Harvey:1990qw}; this scenario is known as Leptogenesis \cite{Fukugita:1986hr}. In general, Leptogenesis scenarios in the Type I and Type III Seesaw Mechanisms are typically at very high energy scales and are subsequently difficult to probe. The Type II Seesaw Mechanism may offer opportunities to connect these high and low scale dynamics. Unfortunately, the Type II Seesaw Mechanism is unable to successfully lead to thermal Leptogenesis without the inclusion of additional particles; an extra triplet Higgs or a right-handed neutrino \cite{Ma:1998dx, Hambye:2005tk}. This is disappointing since the explanation of the neutrino masses only requires a single triplet Higgs, destroying the minimal nature of the model. 

However, since the triplet Higgs is a scalar particle, there are alternative approaches by which the observed baryon asymmetry may be generated. For example, Electroweak Baryogenesis may be possible if the EWPT becomes strongly first-order due to the new interactions introduced through the triplet Higgs, although additional $\mathcal{CP}$ violation sources are also required. Unfortunately, for this scenario to be successful, the triplet Higgs mass must be below $\sim$500 GeV~\cite{Zhou:2022mlz}, which is already in strong tension with current LHC searches.  Another interesting alternative to consider is the Affleck-Dine mechanism~\cite{Affleck:1984fy}. This mechanism requires (i) a scalar that is charged under some mixture of the global $U(1)_L$ or  $U(1)_B$ symmetries; (ii) a small term in the Lagrangian that breaks this symmetry; and (iii) a displaced vacuum value during the early universe. Interestingly, the triplet Higgs carries a $U(1)_L$ charge due to its Yukawa coupling with the leptons, which generate the neutrino masses. Also, $U(1)_L$ breaking terms naturally arise from its couplings with the SM Higgs. However, condition (iii) is less trivially achieved. 

In the Affleck-Dine mechanism, it is assumed that the scalar field exhibits a flat direction, a trajectory where the mass term dominates the potential and the quartic term vanishes, and thus it is easily displaced from the minimum of its potential during the inflationary epoch. For $m_{\textrm{scalar}}\ll H_\textrm{inf}$, the correlation length of the displaced value can be much larger than the current universe and  it may be assumed to take a constant value. Thus, the first and usual implementation of the Affleck-Dine mechanism involves the framework of Supersymmetry, since one can easily find many $D$ and $F$ flat directions for a scalar field that carries $U(1)_L$ or  $U(1)_B$ charge - the superpartners of the SM fermions \cite{Dine:1995uk, Dine:1995kz}. Indeed, the a model of Supersymmetry that includes two triplet superfields was considered in the past as a possible mechanism to generate the baryon asymmetry, with two scalar triplets required to cancel anomalies \cite{Senami:2001qn}. In contrast to these scenarios, we want to consider a minimal framework that does not include Supersymmetry. We emphasize that it is non-trivial to consider the Affleck-Dine mechanism without Supersymmetry since a general model is immune to the presence of flat directions. Additionally, difficulties can arise in finding natural mechanisms to transfer the generated asymmetry from the scalar sector to SM fermions without Supersymmetry. One may instead wonder about the possibility of the vacuum displacement originating from the stochastic behavior of the scalar field during the early inflationary epoch. Since the correlation length of the vacuum value is short $\mathcal O(\frac{1}{H_\textrm{inf}})$ for a quartic coupling of order of $\mathcal{O}(1)$~\cite{Starobinsky:1994bd}, the average of the baryon asymmetry of our universe is approximately zero in this case due to the stochastic nature of the initial phase.

Therefore, the remaining option is to consider the triplet scalar as a component of the inflaton, due to the setup of the inflationary phase requiring an initially displaced vacuum value~\cite{Brout:1977ix, Sato:1980yn, Guth:1980zm, Linde:1981mu, Albrecht:1982mp}. The possibility of the inflaton playing a role in the generation of the baryon asymmetry of the universe has attracted significant attention in recent years~\cite{Hertzberg:2013mba, Lozanov:2014zfa, Yamada:2015xyr, Bamba:2016vjs,Bamba:2018bwl, Cline:2019fxx,Barrie:2020hiu, Lin:2020lmr, Kawasaki:2020xyf, Kusenko:2014lra, Wu:2019ohx, Charng:2008ke, Ferreira:2017ynu, Rodrigues:2020dod, Lee:2020yaj, Enomoto:2020lpf, Mohapatra:2021aig, Mohapatra:2022ngo}. The inflaton candidates that are typically considered are difficult to connect directly to SM physics due to the large discrepancy in their associated energy scales. The inflaton in our setup will consist of the triplet Higgs, in combination with the SM Higgs \cite{Starobinsky:1980te,Whitt:1984pd,Jakubiec:1988ef,Maeda:1988ab,Barrow:1988xh,Faulkner:2006ub,Bezrukov:2011gp}, providing a unique connection between terrestrial experiments and the high energy dynamics of the early Universe. 

In the past, there have been attempts to explain the origins of inflation, Baryogenesis, and the neutrino masses within a simple, unified framework ~\cite{Murayama:1992ua, Murayama:1993xu,Barrie:2020hiu}. However, it is difficult to naturally provide solutions to these unknowns simultaneously with a single addition to the SM. We will present a model that represents a simple and well-motivated realization of this idea.

In our initial study~\cite{Barrie:2021mwi}, we determined that the introduction of the triplet Higgs of the Type II Seesaw Mechanism to the SM is able to explain simultaneously the origins of inflation, Baryogenesis, and the neutrino masses. In this work, we will scrutinise this scenario further, provide the technical details of the model, and discuss its extensive phenomenological implications. The structure of this paper is as follows: In Section \ref{T2SS}, we provide a summary of the Type II Seesaw Mechanism and describe the overall framework of this scenario. Section \ref{Infl} details the inflationary setup and includes a discussion of the induced inflationary trajectory. The Affleck-Dine mechanism for Baryogenesis and its application to the Type II Seesaw Mechanism is explored in detail in Section \ref{ADBaryo}, and the predicted baryon asymmetry is presented. In Section \ref{Phenom}, we examine and combine all of the phenomenological implications of this scenario at colliders and neutrino experiments, as well as the cosmological constraints. Finally, in Section \ref{Conc}, the final results and future avenues for exploration are discussed.


\section{The Type II Seesaw Mechanism and Model Framework}
\label{T2SS}

The Type II Seesaw Mechanism provides a well-motivated and natural framework for explaining the smallness of the observed neutrino masses. This scenario involves the minimal extension of the SM scalar sector by a $SU(2)_L$ triplet scalar $\Delta$ which carries a hypercharge of $1$. The triplet and SM doublet Higgs' are parameterized as follows,
\begin{eqnarray}
H =\left(
\begin{array}{c}
 ~~ h^{+}      \\
  h     
\end{array}
\right)
,~~
\Delta =\left(
\begin{array}{cc}
  \Delta^+/\sqrt{2} & \Delta^{++}      \\
  \Delta^0& -\Delta^+/\sqrt{2}     
\end{array}
\right)~,
\end{eqnarray}
where $h$ and $\Delta^0$ are the neutral components of $H$  and $\Delta$ respectively. In addition to the neutral component, the charged components of the triplet Higgs, $\Delta^+$ and $\Delta^{++}$, play an important role at terrestrial collider experiments. Their phenomenological implications will be integral to the discovery of the Type II Seesaw Mechanism and the determination of its properties \cite{Chongdar:2021tgm, Dev:2021axj, Mandal:2022zmy, Cheng:2022jyi}.

Importantly, the  addition of the triplet Higgs leads to new interactions involving the SM Higgs and the left-handed lepton doublet. In particular, the following Yukawa interaction between the left-handed lepton doublets $L_i$ and the triplet Higgs $\Delta$, 
\begin{equation}
{\mathcal L}_{\textrm{Yukawa}} = \mathcal L^{\rm SM}_{\textrm{Yukawa}}-\frac{1}{2}y_{ij} \bar L^c_i \Delta L_j + h.c. 
\label{nu_interaction1}
\end{equation}

This Yukawa interaction is integral to the generation of a non-zero neutrino mass matrix, which is generated when the neutral component of the triplet Higgs  $\Delta^0$  obtains its non-zero VEV. Additionally, this interaction allows us to assign a lepton charge of $Q_L=-2$ to the triplet Higgs, thus fulfilling one of the conditions for the Affleck-Dine mechanism, and opening the possibility for it to play a role in the origins of the baryon asymmetry via Leptogenesis. 

The inclusion of the triplet Higgs scalar leads to new terms in the Higgs' potential $V(H, \Delta)$, including interactions associated with $\Delta$ that violate the global lepton number symmetry. The potential is given by,
\begin{eqnarray}
V(H, \Delta) &=& -m_H^2 H^\dagger H + \lambda_H  (H^\dagger H)^2 + m_\Delta^2 {\rm Tr}(\Delta^\dagger \Delta) +  \lambda_1 (H^\dagger H)  {\rm Tr}(\Delta^\dagger \Delta) 
+ \lambda_2 ({\rm Tr}(\Delta^\dagger \Delta))^2  \nonumber \\
&& + \lambda_3 {\rm Tr}(\Delta^\dagger \Delta)^2  + \lambda_4  H^\dagger \Delta \Delta^\dagger H  + \left[\mu( H^T i \sigma^2 \Delta^\dagger H) + \frac{\lambda_5}{M_p} ( H^T i \sigma^2 \Delta^\dagger H) (H^\dagger H)\right. \nonumber \\
&& \left. + \frac{\lambda^\prime_5}{M_p} ( H^T i \sigma^2 \Delta^\dagger H) (\Delta^\dagger \Delta) +h.c.\right] +...~,
\label{full_pot1}
\end{eqnarray}
where the terms given in the square brackets $ [...] $ violate the lepton number symmetry. The cubic $\mu$ dependent term is important for determining the VEV obtained by $\Delta^0$. 

In addition to the cubic term, we have included dimension five operators which are suppressed by $M_p$, since these terms may dominate over the $\mu$ term during inflation when the field values are close to the Planck scale. Even if these interaction terms play no role in low energy physics, they could be important during the early universe epochs of inflation and reheating, particularly if the triplet Higgs is a component of the inflaton.

In the cosmological context, our analysis will focus on the neutral components of $\Delta$ and $H$, which have non-trivial VEVs. Thus, the scalar potential $ V(H, \Delta) $ can be simplified as follows,
\begin{eqnarray}
V(h, \Delta^0) &&= -m_H^2 |h|^2+ m_\Delta^2 |\Delta^0|^2 + \lambda_H  |h|^4 + {\lambda_\Delta}  |\Delta^0|^4  + {\lambda_{H\Delta}} |h|^2 |\Delta^0|^2  \nonumber \\
&&  -\left(\mu h^2 {\Delta^0}^* + \frac{\lambda_5}{M_p} |h|^2  h^2  {\Delta^0}^*  + \frac{\lambda^\prime_5}{M_p} |\Delta^0|^2 h^2  {\Delta^0}^* +h.c. \right) +...  ~,
\label{simp_pot1}
\end{eqnarray}
where 
$ \lambda_\Delta = \lambda_2+\lambda_3 $ and  ~$ \lambda_{H\Delta} = \lambda_1+\lambda_4 $. It is important to note that, all of these parameters must have values such that they satisfy the vacuum stability conditions. Also, in the early universe, we require both $h$ and $\Delta^0$ to have non-vanishing vacuum values to ensure that the $U(1)_L$ breaking term is relevant.

From the potential given in Eq. (\ref{simp_pot1}), we can derive the VEV of the triplet Higgs. In the limit where the SM Higgs VEV is much smaller than the $\Delta $ mass parameter,  $m_\Delta \gg v_{\textrm{EW}}$, the non-vanishing $\Delta^0$ VEV is approximately given by,
\begin{equation}
v_\Delta \equiv \langle \Delta^0 \rangle \simeq \frac{\mu v_{\textrm{EW}}^2}{ 2 m^2_\Delta}~,
\end{equation}
where the SM Higgs VEV is $v_{\textrm{EW}} = 246$ GeV. 
The LHC has placed a lower limit on the mass parameter of $m_\Delta \gtrsim 800 $ GeV from searches for the associated doubly-charged Higgs~\cite{ATLAS:2017xqs}. Importantly, for the mass range $m_\Delta > 800$ GeV, the masses of the charged and neutral components of the triplet Higgs are approximately equivalent, $m_{\Delta^{++}} \simeq m_{\Delta^{+}} \simeq m_{\Delta^{0}} \simeq m_\Delta $.

From the Yukawa interaction term in Eq. (\ref{nu_interaction1}), we obtain the mass matrix of the neutrinos,
\begin{equation}
m^\nu_{ij} = y_{ij} v_\Delta~.
\end{equation}

The neutrino masses can be derived by diagonalizing the above matrix by the PMNS matrix. The neutrino Yukawa coupling $y_\nu$ should be smaller than $\mathcal{O}(1)$ to ensure that it remains perturbative up to the Planck scale.

The VEV of $\Delta^0$  has the following range of allowed values,
\begin{equation}
\mathcal{O}(1) \textrm{~GeV}>|\langle \Delta^0 \rangle| \gtrsim 0.05 \textrm{~eV}~,
\end{equation}
where the upper bound on the $\Delta^0$ VEV is derived from the T-parameter constraints determined from precision measurements \cite{Kanemura:2012rs}, and the lower limit ensures the generation of the observed neutrino masses while also requiring perturbative Yukawa couplings.

If the doublet-triplet Higgs model presented above is responsible for inflation, it contains all of the ingredients required to generate a baryon asymmetry during inflation through the Affleck-Dine mechanism. However, the current observational data from Cosmic Microwave Background (CMB) measurements exclude the simple polynomial potential found in Eq. (\ref{simp_pot1}) as the source of the inflationary epoch because it is not sufficiently flat  \cite{Planck:2018jri}. A solution to this problem is the addition of non-minimal couplings between the Higgs' and the Ricci scalar, that is a key feature of standard Higgs inflation, which we shall discuss in detail in the next section.


\section{The Inflationary Setup} 
\label{Infl}

The epoch of Inflation is an established component of standard cosmology due to its ability to solve some of the known observational problems - such as the flatness, horizon, and monopole problems, as well as being able to provide measurable predictions in the form of primordial perturbations \cite{Guth:1980zm,Linde:1981mu,Albrecht:1982wi,Mukhanov:1981xt}. In our scenario, the inflationary setting will be induced by a combination of the SM and triplet Higgs' which have non-minimal couplings to gravity. These couplings act to flatten the scalar potential at large field values, allowing the slow-roll parameters to be sufficiently satisfied and the expected observational signatures to be consistent with current CMB measurements. This form of inflationary mechanism has been well-studied, being the setup for standard Higgs inflation, and results in a Starobinsky-like inflationary epoch \cite{Starobinsky:1980te,Bezrukov:2007ep,Bezrukov:2008ut,GarciaBellido:2008ab,Barbon:2009ya,Barvinsky:2009fy,Bezrukov:2009db,Giudice:2010ka,Bezrukov:2010jz,Burgess:2010zq,Lebedev:2011aq,Lee:2018esk,Choi:2019osi}. 
Combining the non-minimal couplings with the scalar sector discussed in Section \ref{T2SS}, we find the relevant Lagrangian for our model,
\begin{eqnarray}
\frac{\mathcal L}{\sqrt{-g}} & & = -\frac{1}{2} M_p^2 R -F(H,\Delta) R -g^{\mu\nu} (D_\mu H)^\dagger (D_\nu H) \\
&& -g^{\mu\nu} (D_\mu \Delta)^\dagger (D_\nu \Delta) -V(H, \Delta) + \mathcal L_{\textrm{Yukawa}}~,\label{Lagrange1}
\end{eqnarray}
refer to Eq. (\ref{nu_interaction1}) and Eq. (\ref{full_pot1}) for the Yukawa couplings and scalar potential, respectively. 
The non-minimal couplings have the form, 
\begin{equation}
F(H,\Delta) =  \xi_H |h|^2 + \xi_\Delta |\Delta^0|^2=\frac{1}{2}\xi_H \rho^2_H+\frac{1}{2} \xi_\Delta \rho^2_\Delta~,
\label{non-min1}
\end{equation}
where we have introduced the following polar coordinate parametrizations,
\begin{equation}
h \equiv \frac{1}{\sqrt{2}} \rho_{H} e^{i\eta}  ~~\textrm{and}~~ \Delta^0 \equiv \frac{1}{\sqrt{2}}\rho_{\Delta} e^{i\theta}~. 
\label{polar}
\end{equation}

To analyse the inflationary dynamics and observables we must understand the trajectory along which the inflaton evolves. An inflationary setting consisting of two non-minimally coupled scalars  has been found to exhibit a unique inflationary trajectory that is dependent upon the relative size of each scalars' non-minimal and quartic self-couplings \cite{Lebedev:2011aq}. In the large $\xi$ and field limits, the ratio of the two Higgs scalars is fixed as,
\begin{equation}
\frac{ \rho_{H}}{\rho_{\Delta}} \equiv \tan \alpha = \sqrt{\frac{2\lambda_\Delta \xi_H -\lambda_{H\Delta} \xi_\Delta}{ 2\lambda_H \xi_\Delta -\lambda_{H\Delta} \xi_H  }}~.
\label{alpha1}
\end{equation}

To ensure this is the trajectory taken by the inflaton, we require $2\lambda_\Delta \xi_H -\lambda_{H\Delta} \xi_\Delta>0$ and $2\lambda_H \xi_\Delta -\lambda_{H\Delta} \xi_H>0$. Below, we provide  the details of the derivation of this trajectory, following  Ref.~\cite{Lebedev:2011aq}, to demonstrate its existence and properties. 


\subsection{Derivation of the Inflationary Trajectory}
\label{Inf_tra_dec}

Firstly, consider the following field redefinitions of $ h $ and $ \Delta^0 $,
	\begin{equation}
 \chi =  \sqrt{\frac{3}{ 2}} M_p \log \left(1+ \frac{\xi_H |h|^2}{ M_p^2} +\frac{\xi_\Delta (|\Delta^0|)^2}{ M_p^2}\  \right) \textrm{~~and~~} \kappa = \frac{\rho_H}{ \rho_\Delta} ~,
	\end{equation}
using the polar coordinate parametrization given in Eq. (\ref{polar}). Here we will temporarily ignore the motion of the angular directions, $\eta$ and $\theta$, since the potential is dominated by the dynamics of the radial directions, $\rho_H$ and $\rho_\Delta$.
Under this redefinition, the kinetic terms of the Lagrangian become, 
\begin{align}
{\cal L}_{\rm kin} =& \frac{1}{ 2} \biggl( 1+ \frac{1}{ 6} \frac{\kappa^2 +1}{
	\xi_H \kappa^2 +\xi_\Delta} \biggr)~ (\partial_\mu \chi)^2 + 
\frac{M_p}{\sqrt{6}}  ~\frac{(\xi_\Delta-\xi_H) \kappa }{ (\xi_H \kappa^2 +\xi_\Delta)^2} 
(\partial_\mu \chi) (\partial^\mu \kappa) \nonumber\\
&+\frac{M_p^2 }{ 2}\frac{ \xi_H^2 \kappa^2 +\xi_\Delta^2 }{ (\xi_H \kappa^2 +\xi_\Delta)^3 }(\partial_\mu \kappa)^2
~.
\end{align}

In our analysis, we require that the non-minimal couplings are large, 
$ \xi_H  $ and $ \xi_\Delta \gg 1$, such that at leading order in $1/\xi$ the kinetic terms are given by,
\begin{equation}
{\cal L}_{\rm kin}= \frac{1}{ 2 } (\partial_\mu \chi)^2 +
\frac{M_p^2}{ 2} \frac{ \xi_H^2 \kappa^2 +\xi_\Delta^2 }{ (\xi_H \kappa^2 +\xi_\Delta)^3 }(\partial_\mu \kappa)^2 ~. 
\end{equation}

 For large $ \xi $ parameters, the mixing term $(\partial_\mu \chi) (\partial^\mu \kappa) $ is suppressed, resulting in a canonically normalised $ \chi $ and a suppression of the kinetic term of $ \kappa $. Below we show the three main regimes for $ \kappa $, that each have a corresponding canonically normalized variable $\kappa'$,
\begin{eqnarray}
&& \xi_\Delta \gg \xi_H  ~~{\rm or}~~ \kappa \rightarrow 0 ~~,~~~~~ \kappa'= \frac{\kappa}{
	\sqrt{\xi_\Delta}}  ~, \nonumber\\
&& \xi_H \gg \xi_\Delta  ~~{\rm or}~~ \kappa \rightarrow \infty ~~,~~~ \kappa'= \frac{1}{
	\sqrt{\xi_H} \kappa}  ~, \nonumber\\
&& \xi_H = \xi_\Delta  ~~,~~~~~~~~~~~~~~~~~~~~  \kappa'= \frac{1}{{\sqrt{\xi_H}}} \arctan \kappa ~.
\end{eqnarray}

Now consider the scalar potential under this field redefinition, in terms of $ \kappa $,
\begin{equation}
U= \frac{\lambda_H \kappa^4 + \lambda_{h\Delta} \kappa^2 +\lambda_\Delta  }{
	4 (\xi_H \kappa^2 +\xi_\Delta)^2}  M_p^4~,
\end{equation}
where we have taken the large field limit for $ \chi $ and assume that the $U(1)_L$ breaking terms can be ignored. This potential has the following minima, dependent upon the relations between the different non-minimal couplings and quartic self-couplings,
\begin{align}
 (1)&~2 \lambda_H \xi_\Delta - \lambda_{h\Delta} \xi_H >0~,~
2 \lambda_\Delta \xi_H - \lambda_{h\Delta} \xi_\Delta >0~,~~~~\kappa = \sqrt{\frac{ 
	2 \lambda_\Delta \xi_H - \lambda_{h\Delta} \xi_\Delta  }{ 
	2 \lambda_H \xi_\Delta - \lambda_{h\Delta} \xi_H}   }  ~,  \nonumber\\
 (2)&~2 \lambda_H \xi_\Delta - \lambda_{h\Delta} \xi_H >0~,~
2 \lambda_\Delta \xi_H - \lambda_{h\Delta} \xi_\Delta <0~,~~~~\kappa=0  ~,  \nonumber\\
(3)&~2 \lambda_H \xi_\Delta - \lambda_{h\Delta} \xi_H <0~,~
2 \lambda_\Delta \xi_H - \lambda_{h\Delta} \xi_\Delta >0~,~~~~\kappa=\infty  ~, \nonumber\\
 (4)&~2 \lambda_H \xi_\Delta - \lambda_{h\Delta} \xi_H <0~,~
2 \lambda_\Delta \xi_H - \lambda_{h\Delta} \xi_\Delta <0~,~~~~\kappa=0,\infty  ~. 
\end{align}

In Cases 2 and 3, we arrive at the usual single field non-minimally coupled inflationary  potential,  $\lambda_\Delta /(4\xi_\Delta^2) M_p^4$ and 
$\lambda_H /(4\xi_H^2) M_p^4$, respectively. The inflaton in Case 1 is characterised by a mixture of the two scalars which is the typical scenario we require to ensure that the lepton number violating interactions are relevant during inflation.  Importantly in Case 1, the potential has the following minimum,
\begin{equation}
U\Bigl\vert_{\rm min~(1) }= \frac{1}{16} \frac{4 \lambda_\Delta \lambda_H - \lambda_{h\Delta}^2}{\lambda_\Delta \xi_H^2 + \lambda_H \xi_\Delta^2 - \lambda_{h\Delta} \xi_\Delta \xi_H } M_p^4~.
\label{Umin1}
\end{equation}

The numerator of Eq. (\ref{Umin1}) must be positive to ensure that we do not obtain a negative vacuum energy at large field values. In each of the above cases, the canonical field $\kappa'$ has a large mass which is of order  $M_p/ \sqrt{\xi}$. This mass will always be greater than the Hubble parameter during inflation, so we can safely integrate out the $ \kappa $ field, establishing the inflationary trajectory. 

Having derived this inflationary trajectory, we can now proceed to analyzing the inflaton dynamics in our Lagrangian given in Eq. (\ref{Lagrange1}). Since we only wish to consider Case 1, we can simplify our analysis by defining the inflaton as $ \varphi $, through the following relations to the polar coordinate fields,
\begin{eqnarray}
&&  \rho_{H} = \varphi \sin \alpha,~\rho_{\Delta} = \varphi \cos \alpha~,  \nonumber \\
&& \xi \equiv \xi_H \sin^2 \alpha + \xi_\Delta \cos^2 \alpha~.
\end{eqnarray}

The Lagrangian in Eq. (\ref{Lagrange1}) is now given by,
\begin{equation}
\frac{\mathcal L}{\sqrt{-g}} = -\frac{M_p^2}{2}  R -\frac{\xi}{2}   \varphi^2  R  - \frac{1}{2} g^{\mu\nu} \partial_\mu \varphi \partial_\nu \varphi  -\frac{1}{2} \varphi^2 \cos^2\alpha~ g^{\mu\nu} \partial_\mu \theta \partial_\nu \theta-V(\varphi, \theta) ~,
\label{lagrang1}
\end{equation}
where 
\begin{equation}
\hspace{-0.5cm}V(\varphi, \theta)  = \frac{1}{2} m^2 \varphi^2 + \frac{\lambda}{4} \varphi^4 + 2\varphi^3 \left(\tilde \mu   + \frac{\tilde \lambda_5}{M_p} \varphi^2\right) \cos\theta ~,
\end{equation}
and
\begin{eqnarray}
 && m^2 = m^2_{\Delta} \cos^2 \alpha  -m_H^2 \sin^2 \alpha~,  \nonumber \\
 && \lambda = \lambda_H \sin^4 \alpha + \lambda_{H\Delta} \sin^2 \alpha \cos^2 \alpha + \lambda_{\Delta} \cos^4 \alpha~,  \nonumber \\
 && \tilde \mu  = -\frac{1}{2\sqrt{2}}\mu \sin^2\alpha  \cos \alpha~,  \nonumber \\
 && \tilde \lambda_5  = - \frac{1}{4\sqrt{2}} (\lambda_5 \sin^4\alpha  \cos \alpha + \lambda^\prime_5 \sin^2\alpha  \cos^3 \alpha)~.
\end{eqnarray}

The Affleck-Dine mechanism will be realised in our scenario through the motion of the dynamical field $\theta$. The size of the generated lepton asymmetry will be determined by the size of the non-trivial motion induced in $\theta$ sourced by inflation. During inflation, $m \ll \varphi$, which means that the quartic term in the Jordan frame potential dominates the inflationary dynamics. 


\subsection{The Single Field Inflation Approximation}
\label{singlefield}
 We will choose input parameters that preserve the inflationary trajectory and ensure that it is negligibly affected by the dynamics of $ \theta $ and the lepton asymmetry production. In this case, the inflationary observables in our framework are consistent with those of the Starobinsky inflationary model, and as such, are in excellent agreement with current observational constraints \cite{Planck:2018jri}.
 
Since we have derived the approximate single field inflationary behaviour above, we can analyse the evolution of the inflationary phase and its predictions for the inflationary parameters. To do this, we will  first  translate the Lagrangian in Eq. (\ref{lagrang1}) from the Jordan frame to the Einstein frame, using the following transformations \cite{Wald:1984rg,Faraoni:1998qx}, 
\begin{equation}
\tilde{g}_{\mu \nu} =\Omega^2 g_{\mu\nu},~~ \Omega^2= 1+ \xi \frac{\varphi^2}{M^2_p}~,
\end{equation}
and subsequently reparametrizing $\varphi$ in terms of the canonically normalized scalar $\chi$.

 In transforming from the Jordan to Einstein frame, the Jordan frame field $ \varphi $ no longer has a canonically normalised kinetic term. We must make the following field redefinition to obtain the canonically normalized scalar,
\begin{equation}
\frac{d\chi}{d\varphi}
\;=\; \dfrac{\sqrt{(6\xi^2 \varphi^2/M_p^2)+\Omega^2}}{\Omega^2}
\;~,
\end{equation}
which gives the definition of $\chi$ in terms of $ \varphi $ as, 
\begin{equation}
\chi(\varphi) = \frac{1}{\sqrt{\xi}} \left(\sqrt{1+6\xi}~ \sinh^{-1}(\sqrt{\xi+6\xi^2}  \varphi)  -\sqrt{6\xi} ~ \sinh^{-1} (\sqrt{6\xi^2}\varphi/\sqrt{1+ \xi \varphi^2})\right)~.
\end{equation} 

Three distinct regimes can be observed in the evolution of the relation between the $ \chi $ and $ \varphi $ fields, namely,
 \begin{equation}
 \dfrac{\chi}{M_p}
 \approx
 \left\{
 \begin{array}{lll}
 \dfrac{\varphi}{M_p}
 & \mbox{for $\dfrac{\varphi}{M_p} \ll\dfrac{1}{\xi}$} 
 & \mbox{(radiation-like)}
 \\
 \sqrt{\dfrac{3}{2}}\,\xi\left(\dfrac{\varphi}{M_p}\right)^2 
 & \mbox{for $\dfrac{1}{\xi}\ll \dfrac{\varphi}{M_p} \ll \dfrac{1}{\sqrt{\xi}}$\quad} 
 & \mbox{(matter-like)}
 \\
 \sqrt{\dfrac{3}{2}} \ln\Omega^2 =
 \sqrt{\dfrac{3}{2}} \ln\left[1+\xi\left(\dfrac{\varphi}{M_p}\right)^2\right] \quad
 & \mbox{for $\dfrac{1}{\sqrt{\xi}}\ll \dfrac{\varphi}{M_p}$} 
 & \mbox{(inflation)}
 \end{array}
 \right.
 \label{chi-varphi-relation_sm1}
 \end{equation}

Thus, we now obtain the final Einstein frame Lagrangian,
\begin{eqnarray}
\frac{\mathcal L}{\sqrt{-g}} = -\frac{M_p^2}{2}  R   - \frac{1}{2}  g^{\mu\nu} \partial_\mu \chi \partial_\nu \chi -
 \frac{1}{2} f(\chi)  g^{\mu\nu} \partial_\mu \theta \partial_\nu \theta- U(\chi,\theta) ~,
\label{lagrangian2}
\end{eqnarray}
where 
\begin{eqnarray}
&& \hspace{-0.5cm}f(\chi) \equiv  \frac{\varphi(\chi)^2 \cos^2 \alpha}{\Omega^2(\chi)},  ~~\textrm{and}~~ U(\chi,\theta) \equiv  \frac{V(\varphi(\chi), \theta)}{\Omega^4(\chi)}~.
\label{potential1}
\end{eqnarray}

As with the relation between the $ \chi $ and $ \varphi $ fields, the scalar potential exhibits three distinct regimes in each field definition. Importantly, it is evident that the non-minimal coupling leads to a flattening of the potential in the large field limit. The scalar potential in each regime is,
\begin{equation}
U(\chi) \approx
\left\{
\begin{array}{lll}
\dfrac{1}{4}\lambda\chi^4 
& \mbox{for $\dfrac{\chi}{M_p}\ll\dfrac{1}{\xi}$}
& \mbox{(radiation-like)}
\\
\dfrac{1}{2} m_{S}^2 \chi^2 
& \mbox{for $\dfrac{1}{\xi} \ll \dfrac{\chi}{M_p} \ll 1$\quad}
& \mbox{(matter-like)}
\\
\dfrac{3}{4}m_{S}^2 M_p^2
\left(1-e^{-\sqrt{\frac{2}{3}}(\chi/M_p)}\right)^2 \quad
& \mbox{for $1\ll \dfrac{\chi}{M_p}$}
& \mbox{(inflation)}
\end{array}
\right.
\label{Uchi_sm1}
\end{equation}

Therefore, the $ \chi $ potential replicates the Starobinsky form in the large field limit,
\begin{equation}
U_{\textrm{inf}}(\chi)= \frac{3}{4}m_S^2 M_p^2 \left(1-e^{-\sqrt{\frac{2}{3}}\frac{\chi}{Mp}}\right)^2~,
\label{inf_pot1}
\end{equation}
where the Starobinsky mass scale is required by observational constraints to be $ m_S =\sqrt{\frac{\lambda M_p^2}{3\xi^2}}\simeq 3 \cdot 10^{13} $ GeV  \cite{Faulkner:2006ub,Planck:2018jri}. This will be the inflationary framework we will use in our scenario, and from which we will now determine the expected observational signatures. 

\begin{figure}
      	\centering
      	\includegraphics[width=0.7\textwidth]{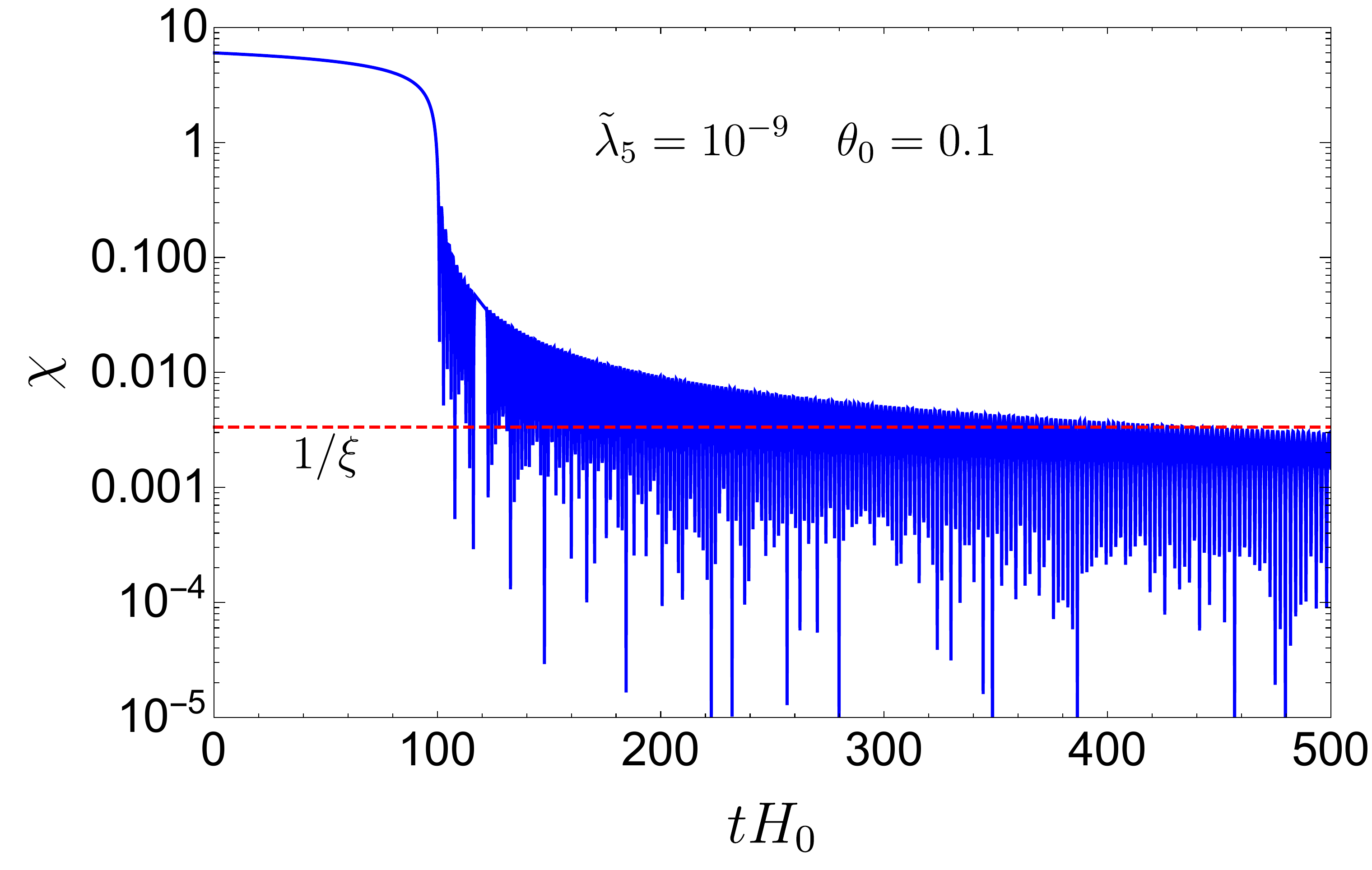}
      	\caption{The evolution of $\chi$ during and after inflation is depicted, in Planck units. The following input parameters are chosen to be $ \xi= 300$ and $\lambda=4.5 \cdot 10^{-5}$, with the initial conditions  $\chi_0=6.0 M_p$, $\dot \chi_0=0$,  and $\dot \theta_0=0 $ used. We require that $\tilde\mu$ is sufficiently small to not affect the inflationary dynamics, and we define $H_0\equiv m_S/2$. }
      	\label{chi1}
      \end{figure}

Before calculating the inflationary observables, we provide a plot in Figure \ref{chi1} of the evolution of $\chi$ during and after inflation, where we have numerically solved the equations of motion derived from the Lagrangian in Eq. (\ref{lagrangian2}). The input parameters are fixed to $ \xi= 300$ and  $\lambda=4.5 \cdot 10^{-5}$, we take $\tilde\mu$ to be sufficiently small to not affect the dynamics, and the following initial conditions are chosen, $\dot \theta_0=0 $, $\chi_0=6.0 M_p$, and $\dot \chi_0=0$  with the definition $H_0\equiv m_S/2$. The figure shows that the end of inflation occurs near $tH_0=100$, and thus, horizon crossing occurs at approximately $tH_0 \approx 40-50$. After inflation, the universe quickly enters into a matter-like evolution until $t H_0 \approx 400$. The line  $\chi =  M_p/\xi$ in Figure \ref{chi1} is included to indicate the end of the matter-like epoch. 

It is important to note that, the reason we choose $ \xi= 300$ and $\lambda=4.5 \cdot 10^{-5}$ is that for larger $\xi$ values, or equivalently small $\lambda$, issues can arise regarding unitarity during the oscillatory stage that follows inflation in Higgs inflation models \cite{Ema:2016dny, DeCross:2015uza, DeCross:2016cbs, DeCross:2016fdz}. It has been found that this problem can be avoided if these models are UV completed by the addition of an $R^2$ term~\cite{Giudice:2010ka,Gorbunov:2018llf, Ema:2017rqn, Ema:2019fdd, Ema:2020zvg}. 

Another subtlety is that the $\lambda$ coupling will obtain radiative corrections from the gauge couplings and Yukawa couplings of the SM and triplet Higgs'.  A naive estimation of the size of these radiative corrections is given approximately by - gauge couplings: $\frac{g^4}{16\pi^2}$ ($\mathcal{O}(10^{-(3-4)})$); neutrino Yukawa couplings: $-\frac{y^4}{16\pi^2}$; or the top Yukawa coupling: $-\frac{y_t^4}{16\pi^2}$, for a general mixing angle $\alpha$.  To obtain $\lambda=4.5 \cdot 10^{-5}$, a small tuning of $y$ or $y_t$ might be necessary to cancel the contributions from the Higgs' gauge couplings.  However, we stress that such a small $\lambda$ is not a necessary condition for our mechanism to work, and we will give a comment on our final result for larger couplings. In any case, for couplings of $\lambda \sim 10^{-2}$, the dynamics will be stable against these radiative corrections. Nevertheless, we will use this benchmark coupling for our study to simplify the numerical analysis during the oscillation stage.


\subsection{Inflationary Observables}

Considering the Einstein frame inflationary potential in Eq. (\ref{inf_pot1}), we derive the  following characteristic slow-roll parameters,
\begin{equation}
\epsilon_s\simeq \frac{3}{4N_*^2}~, ~~\textrm{and} ~~\eta_s\simeq \frac{1}{N_*}~,
\label{slowroll_sm1}
\end{equation} 
where we have given them in terms of the number of e-foldings of expansion from horizon crossing to the end of the inflation, $ N_* $. Subsequently, we obtain the predicted spectral index and tensor-to-scalar ratio in our inflationary scenario that should be observed in CMB measurements,
\begin{equation}
n_s\simeq 1-\frac{2}{N_*}~, ~~\textrm{and} ~~r\simeq \frac{12}{N_*^2}~,
\label{inflation_obs_sm1}
\end{equation} 
respectively. The required number of inflationary e-folds to solve the flatness and horizon problems is between $ 50<N_*<60 $. Thus, our model gives the following predicted ranges for the spectral index and tensor-to-scalar ratio,
\begin{eqnarray}
\label{nspre1}
0.96\lesssim n_s \lesssim 0.9667~,\\
\label{rpre1}
 0.0033\lesssim r\lesssim 0.0048~.
\end{eqnarray}

We can now compare these to the current best constraints on the inflationary predictions from the recent Planck results \cite{Planck:2018jri},
\begin{eqnarray}
\label{ns1}
n_s = 0.9649\pm 0.042~~{\rm (68\% C.L.)}~,\\
\label{r1}
r_{0.002} < 0.056~~{\rm (95\% C.L.)}~,
\end{eqnarray}
for the $\Lambda-$CDM$+r$ model. Comparing these with those in Eq. (\ref{nspre1}) and (\ref{rpre1}), we see that our scenario is in extremely good agreement with current observational results for the required number of inflationary e-folds. The prediction for the tensor-to-scalar ratio is well below the current constraints, but it will be observable by the upcoming LiteBIRD telescope \cite{Hazumi:2019lys}. 

In addition to these results, the scalar perturbations observed in the CMB provide a constraint on the Starobinsky mass scale $ m_S $. It is necessary to satisfy the following relation for consistency with this observation,
\begin{equation}
\frac{\lambda}{\xi^2}\simeq 5\cdot 10^{-10} ~,
\end{equation} 
with the corresponding Starobinsky mass scale being $ m_S\simeq 3 \cdot 10^{13} $ GeV. This relation shall inform our parameter choices in the upcoming sections, and was used to make the parameter choices in Figure \ref{chi1}.


\section{The Affleck-Dine Mechanism and Generated Baryon Asymmetry} 
\label{ADBaryo}
The origin of the  baryon asymmetry of the universe is a fundamental mystery of modern  physics.  The size of the observed baryon asymmetry is parametrized by the asymmetry parameter $\eta_{B}$ \cite{Aghanim:2018eyx},  
\begin{equation}
\eta_B^\textrm{obs} = \frac{n_B}{s} \simeq 8.5\cdot 10^{-11} ~,
\end{equation}
where $n_B$ and $s$ are the baryon number and entropy densities of the universe, respectively. Many models have been proposed to explain this asymmetry, with one of the most well-studied scenarios being the Affleck-Dine mechanism \cite{Affleck:1984fy}.

The core idea behind the Affleck-Dine mechanism is the generation of non-zero angular motion in the phase of a complex scalar field $\phi$ that is charged under a global $U(1)$ symmetry. Assuming the complex scalar acquires a large initial field value in the early universe, $ \phi $ will start to oscillate once the Hubble parameter decreases below the $\phi$ mass $m$.  If the scalar potential $ V(\phi) $ contains an explicit $U(1)$ breaking term, a net $U(1)$ charge asymmetry will be generated by this motion. 
Consequently, if the $U(1)$ symmetry is composed of the global $U(1)_B$ or $U(1)_L$ symmetries, a baryon asymmetry can be generated prior to the EWPT. In the following, we will show that the role of the $\phi$ field can be taken by our inflaton, the mixed state of the SM and triplet Higgs' discussed in the previous section, which carries a complex phase associated with the global $U(1)_L$ symmetry.

If we identify this scalar with our inflaton and the complex phase of the triplet Higgs, the asymmetry number density associated with the $U(1)_L$ charge will be given by,
\begin{equation}
n_L=Q_L \varphi^2(\chi) \dot\theta \cos^2 \alpha    ~.
\end{equation}
where $\alpha$ is the mixing angle between the SM and triplet Higgs' during inflation defined in the previous section. Therefore, in order to obtain a non-zero lepton number asymmetry density $n_L$, we require a non-zero vacuum value for $\chi$ and non-trivial motion in the complex phase $\theta$. This net lepton asymmetry shall be transferred to the baryonic sector through equilibrium sphaleron processes prior to the EWPT.


\subsection{Asymmetry Generation from the Triplet Higgs} %

To begin our analysis, consider the  Lagrangian of our model in the Einstein frame, as given in Eq.~(\ref{lagrangian2}),
\begin{equation}
\frac{\mathcal L}{\sqrt{-g}} = -\frac{M_p^2}{2}  R   - \frac{1}{2}  g^{\mu\nu} \partial_\mu \chi \partial_\nu \chi -
 \frac{1}{2} f(\chi)  g^{\mu\nu} \partial_\mu \theta \partial_\nu \theta- U(\chi,\theta) ~,
\end{equation}
where 
\begin{equation}
f(\chi) \equiv  \frac{\varphi(\chi)^2 \cos^2 \alpha}{\Omega^2(\chi)}~,
\label{fchi1}
\end{equation}
and
\begin{equation}
 U(\chi,\theta) \equiv  \frac{V(\varphi(\chi), \theta)}{\Omega^4(\chi)}~=   \frac{m^2 \varphi^2(\chi)+ \lambda \varphi^4(\chi)  }{\Omega^4(\chi)}  +  \frac{2 \tilde \mu \varphi^3(\chi) + 2\frac{\tilde \lambda_5}{M_p}   \varphi^5(\chi)  }{\Omega^4(\chi)} \cos \theta~.
\end{equation}
The equations of motion for $\chi$ and $\theta$ can be simply derived as follows,
\begin{eqnarray}
&& \ddot{\chi} -\frac{1}{2} f_{,\chi}  \dot \theta^2 + 3 H \dot \chi + U_{,\chi} =0  ~,\nonumber \\
&& \ddot{\theta} + \frac{f_{,\chi}}{f(\chi)}  \dot \theta \dot \chi + 3 H \dot \theta +\frac{1}{f(\chi)} U_{,\theta} =0 ~.
\label{thetaEOMa}
\end{eqnarray} 

Evaluating these equations of motion during the inflationary epoch, it is found that both $\chi$ and $\theta$ enter approximate slow-roll regimes described by,
\begin{equation}
\dot{\chi} \simeq -\frac{M_p U_{,\chi}}{\sqrt{3U}}, ~~\textrm{and}~~ \dot{\theta} \simeq -\frac{M_p U_{,\theta}}{f(\chi)\sqrt{3U}} ~.
\label{dottheta_inf1}
\end{equation}

To simplify our analysis,  we make use of the following parameterization  $\tau= t H_0$, where $H_0= m_S/2$ and define prime as the derivative with respect to $\tau$. We also define the reduced Hubble parameter $\tilde H$ by,
\begin{equation}
\tilde{H}^2= \frac{1}{3}\left (\frac{1}{2}{\chi^\prime}^2 + \frac{1}{2} f(\chi) {\theta^\prime}^2 + U/H_0^2 \right ).
\end{equation} 

Under these redefinitions, the equations of motion from Eq. (\ref{thetaEOMa}) become,
\begin{eqnarray}
&& \chi^{\prime\prime} -\frac{1}{2} f_{,\chi}  {\theta^\prime}^2 + 3 \tilde H \chi^\prime + \frac{U_{,\chi}}{H_0^2} =0  ~,\nonumber \\
&& \theta^{\prime\prime} + \frac{f_{,\chi}}{f(\chi)} \theta^\prime \chi^\prime + 3 \tilde H \theta^\prime +\frac{1}{f(\chi)H_0^2} U_{,\theta} =0 ~,
\label{thetaEOM2}
\end{eqnarray} 
and the associated lepton number density is,
\begin{equation}
\dot \theta = \theta^\prime H_0, ~{\rm and} ~ n_L=Q_L \varphi^2(\chi) \theta^\prime \cos^2 \alpha H_0   ~.
\end{equation} 

 \begin{figure}
      \begin{subfigure}
      	\centering
      	\includegraphics[width=0.505\textwidth]{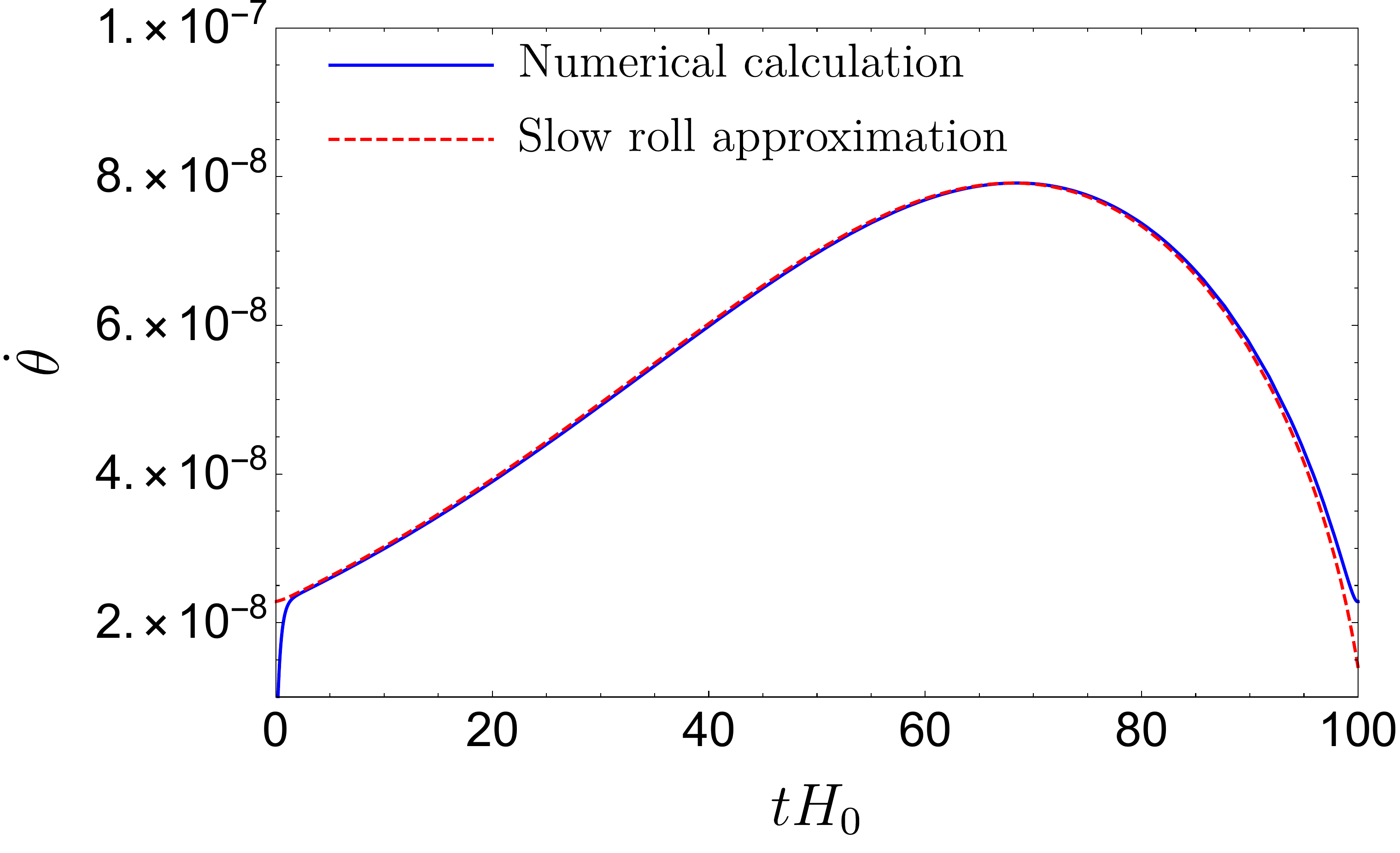}
      	\end{subfigure}
      	\hfill 
      	   \begin{subfigure}
      	   	\centering
      	   	\includegraphics[width=0.475\textwidth]{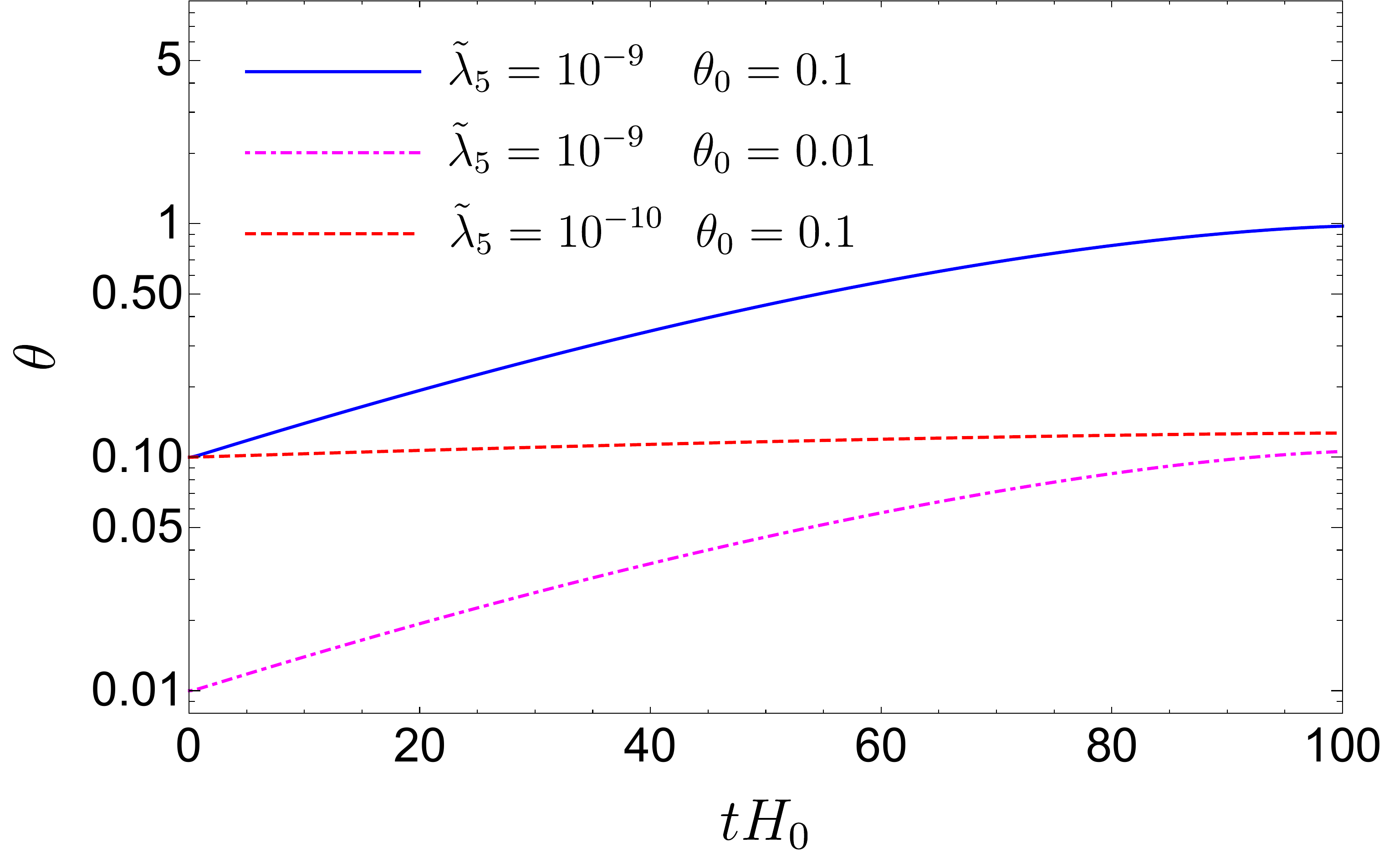}
      	   	\end{subfigure}
      	\caption{  The dynamics of (left) $\dot \theta$ and (right) $\theta$ during inflation are depicted, in Planck units. The left figure provides a comparison of the exact numerical result with the analytical result found in Eq. (\ref{thetaEOMa}) for $\tilde \lambda_5= 10^{-9}$. The other input parameters selected are $ \xi= 300$ and $\lambda=4.5\cdot 10^{-5}$, with the following initial conditions  $\chi_0=6.0 M_p$, $\dot \chi_0=0$, and $\dot \theta_0=0 $ used. We choose $\tilde\mu$ to be sufficiently small to not affect the dynamics, and where we define $H_0= m_S/2$. }
      	\label{theta1}
      \end{figure}

In Figure~\ref{theta1}, we depict the evolution of $\theta$ and $\dot \theta$ during inflation, providing a comparison with the estimation of  $\dot \theta$ presented in Eq.~(\ref{dottheta_inf1}). We find that $\dot \theta$ is well described by the derived slow-roll approximation. Due to the non-zero $\dot \theta$ induced, there is an increase in $\theta$ during inflation. However, as shown in the right panel of Figure~\ref{theta1}, for fixed $\theta_0$, the rate of increase is proportional to $\tilde \lambda_5$, which is indicated by the parameter dependencies found in Eq.~(\ref{dottheta_inf1}).

During inflation, the $\chi$ field approaches $M_{p}$ and so we can ignore the subdominant $m$ and $\tilde{\mu}$ terms that are much less than $ M_{p}$. Using these approximations, we can determine the lepton number asymmetry generated by the motion of the triplet Higgs phase $\theta$ at the end of inflation,
\begin{eqnarray}
{n_L}_\textrm{end}&=& Q_L \varphi^2_\textrm{end} \dot\theta_\textrm{end} \cos^2 \alpha  \nonumber \\
&\simeq& - \mathcal{O}(1) Q_L \varphi^2_\textrm{end} \frac{M_p U_{,\theta}}{f(\chi_{\rm end })\sqrt{3U_{\rm end}}} \cos^2 \alpha  \nonumber \\
&\simeq& - \mathcal{O}(1) Q_L \tilde \lambda_5 \varphi^3_\textrm{end} \sin \theta_{\rm end} /\sqrt{3 \lambda}   ~,
\label{nlend1}
\end{eqnarray}
where the $\mathcal{O}(1)$ factor that appears in the second step is typically $\sim 3$ from numerical calculations, and accounts for the breakdown of the slow-roll approximation at the end of inflation,  which we have defined as when the slow-roll parameter satisfies $\epsilon=1$. In the last step, we assume the quartic term dominates the inflationary potential and that the $\tilde \lambda_5$ coupling dominates the lepton number violating interactions. After inflation, the lepton number density is solely red-shifted by $a^3$ alongside another $\mathcal{O}(1)$ factor given by $\Omega$.


\subsection{The Lepton Number Density Evolution after Inflation} 

When inflation ends, at $\chi_\textrm{end} \simeq 0.67 M_p $, the oscillatory epoch begins for which the universe is characterized by an approximately matter-like evolution. The inflationary potential at this time is given in Eq. (\ref{Uchi_sm1}). At the end of the inflationary stage, we define the lepton asymmetry number density as,
\begin{equation}
{n_L}_\textrm{end}= Q_L \varphi^2_\textrm{end} \dot\theta_\textrm{end} \cos^2 \alpha    ~.
\end{equation}

Now we wish to analyse the evolution of the lepton number density after inflation. To determine this, first consider the following relation derived  from the equation of motion in Eq. (\ref{thetaEOMa}), 
\begin{equation}
\frac{d [  a^3 f(\chi) \dot \theta]}{dt} = \frac{d [ a^3 n_L/(Q_L\Omega^2)]}{dt}  = a^3 U_{,\theta}~.
\label{dlepton1}
\end{equation}

During the matter-like phase, if the term $ U_{,\theta}$ red-shifts faster than matter-like, $a^{-3}$, then it can be ignored and the quantity $a^3 n_L/(Q_L\Omega^2)$ will be conserved. This is guaranteed when the dominant $U(1)_L$ breaking term in $V(\varphi,\theta)$ is a polynomial function of $\varphi$ larger than degree four because the quartic term of $\varphi$ in $V(\varphi,\theta)$ is equivalent to the mass term of the $\chi$ field, which red-shifts as $a^{3}$ after inflation. Thus, we consider the quintic $\tilde{\lambda}_5$ term to be the dominant $U(1)_L$ breaking term at large field values. 

If instead, the cubic term was to dominate,  the $U(1)_L$ breaking term would become increasingly relevant as the expansion of the universe proceeds, and subsequently destroy the lepton number generated during inflation. In the following subsections, we will show that the cubic term is already required to be small to avoid the washout  of the lepton asymmetry after reheating.

As shown in Eq. (\ref{dlepton1}), after inflation the lepton asymmetry number density is red-shifted by the usual matter-like scale factor dependence, with the addition of a factor of $1/\Omega^2$ which is of order of $\mathcal{O}(1)$. It is then possible to estimate the lepton asymmetry number density at any time after inflation using,
\begin{equation}
n_L(t) = {n_L}_\textrm{end} \frac{\Omega^2(\chi)}{\Omega^2(\chi_\textrm{end})} \left (\frac{a}{a_\textrm{end}}\right)^{-3}~.
\label{lepton1}
\end{equation}

\begin{figure}
      	\centering
      	\includegraphics[width=0.7\textwidth]{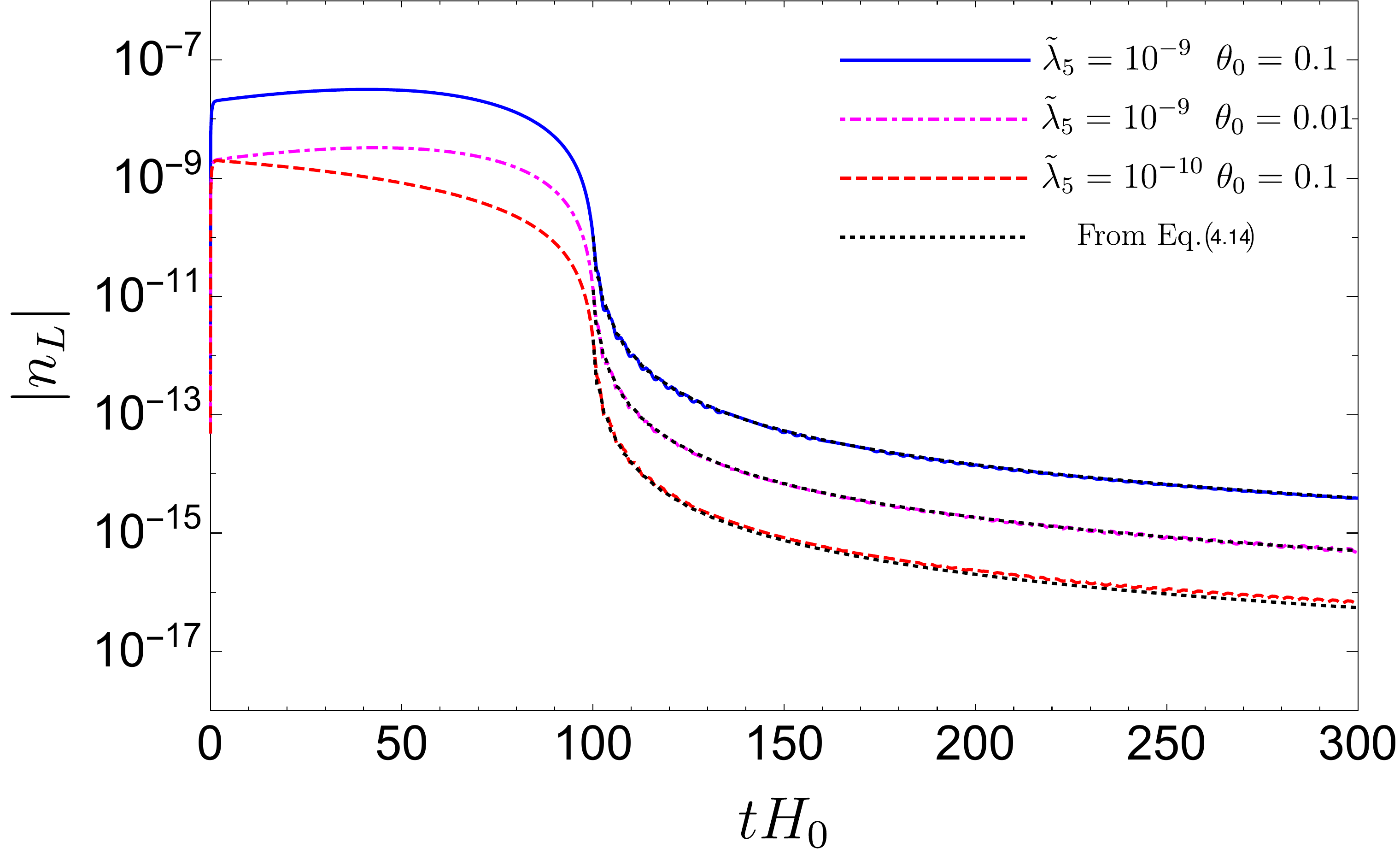}
      	\caption{Comparison of the lepton number density $ n_L $ derived from the full numerical calculations (Coloured lines) and the analytical estimation (Black dotted lines) in Eq. (\ref{lepton1}), through the inflationary and oscillatory epochs. The results are presented for varying $\theta_0$ and $\tilde{\lambda}_5$, with all parameters are in Planck units and we define $H_0= m_S/2$. The fixed input parameters are $ \xi= 300$ and $\lambda=4.5 \cdot 10^{-5}$. We take $\tilde\mu$ to be sufficiently small to not affect the dynamics, and the following initial conditions are chosen, $\chi_0=6.0 M_p$, $\dot \chi_0=0$,  and $\dot \theta_0=0$. }
      	\label{numerical1}
      \end{figure}

In Figure~\ref{numerical1}, we depict a  comparison of the numerical simulations to the analytical result given in Eq. (\ref{lepton1}) to demonstrate the accuracy of this relation. The input parameters have been chosen to fit the current CMB constraints.

In the preceding analysis, it has been assumed that the mixing angle $\alpha$ is fixed throughout the inflation and oscillation epochs. It is observed that the $\Omega^2$ factor quickly approaches 1 after inflation, and thus, the scalar potential can be approximated by $V(h,\Delta^0)$ in Eq. (\ref{simp_pot1}). This potential also exhibits a characteristic minimum direction which is defined by,
\begin{equation}
\frac{ \rho_{H}}{\rho_{\Delta}} \equiv \tan \beta = \sqrt{\frac{2\lambda_\Delta -\lambda_{H\Delta} }{ 2\lambda_H -\lambda_{H\Delta}  }}~,
\end{equation}
where we require $2\lambda_\Delta -\lambda_{H\Delta}>0$ and $2\lambda_H -\lambda_{H\Delta}>0 $ for stability. By comparing this relation to Eq. (\ref{alpha1}), we see that if  $\xi_H = \xi_\Delta$, then the two mixing angles $\alpha$ and $\beta$ converge, and we can  use the same mixing angle to describe the inflationary  and oscillation epochs. If instead, we have $\xi_H \neq \xi_\Delta$, the $\alpha$ and $\beta$ are generally different and a detailed understanding of the evolution during the oscillation stage after inflation requires a dedicated analysis. However, we believe that the lepton asymmetry would only be negligibly affected if this is the case because almost all of the lepton asymmetry is generated during inflation, with its evolution just determined by red-shifting due to the expansion of the universe in the oscillatory phase. It should be noted that the detail of the oscillation epoch, and the transition between unequal $\alpha$ and $\beta$, might affect the expected preheating dynamics and the reheating temperature. The analysis of these dependencies is beyond the scope of this work, and so, we adopt  $\xi_H = \xi_\Delta$ for simplicity.


\subsection{Behaviour of the \texorpdfstring{$\mu$}{text} Term}

\begin{figure}
      \begin{subfigure}
      	\centering
      	\includegraphics[width=0.491\textwidth]{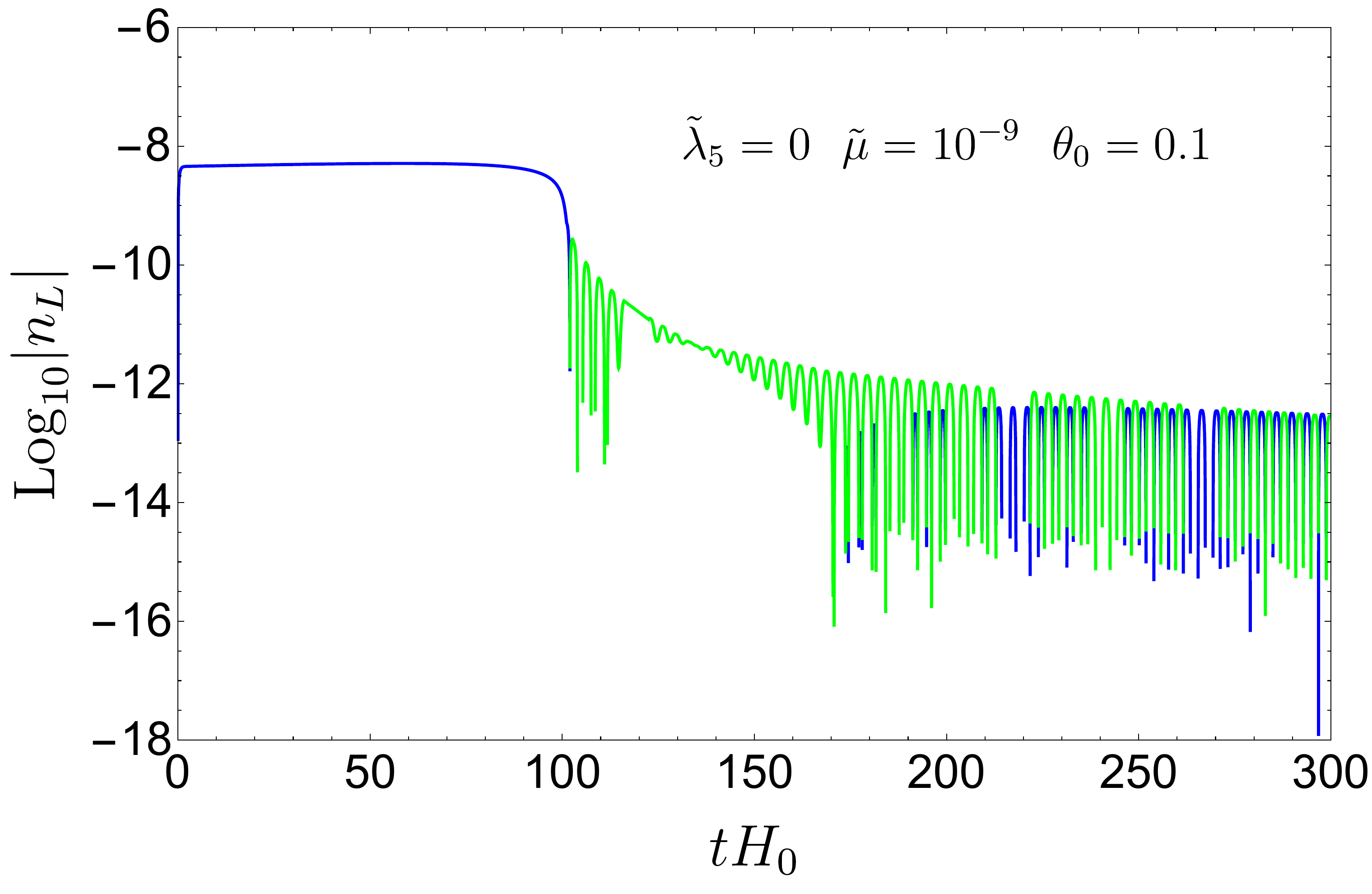}
      	\end{subfigure}
      	\hfill 
      	   \begin{subfigure}
      	  	\centering
      	   	\includegraphics[width=0.491\textwidth]{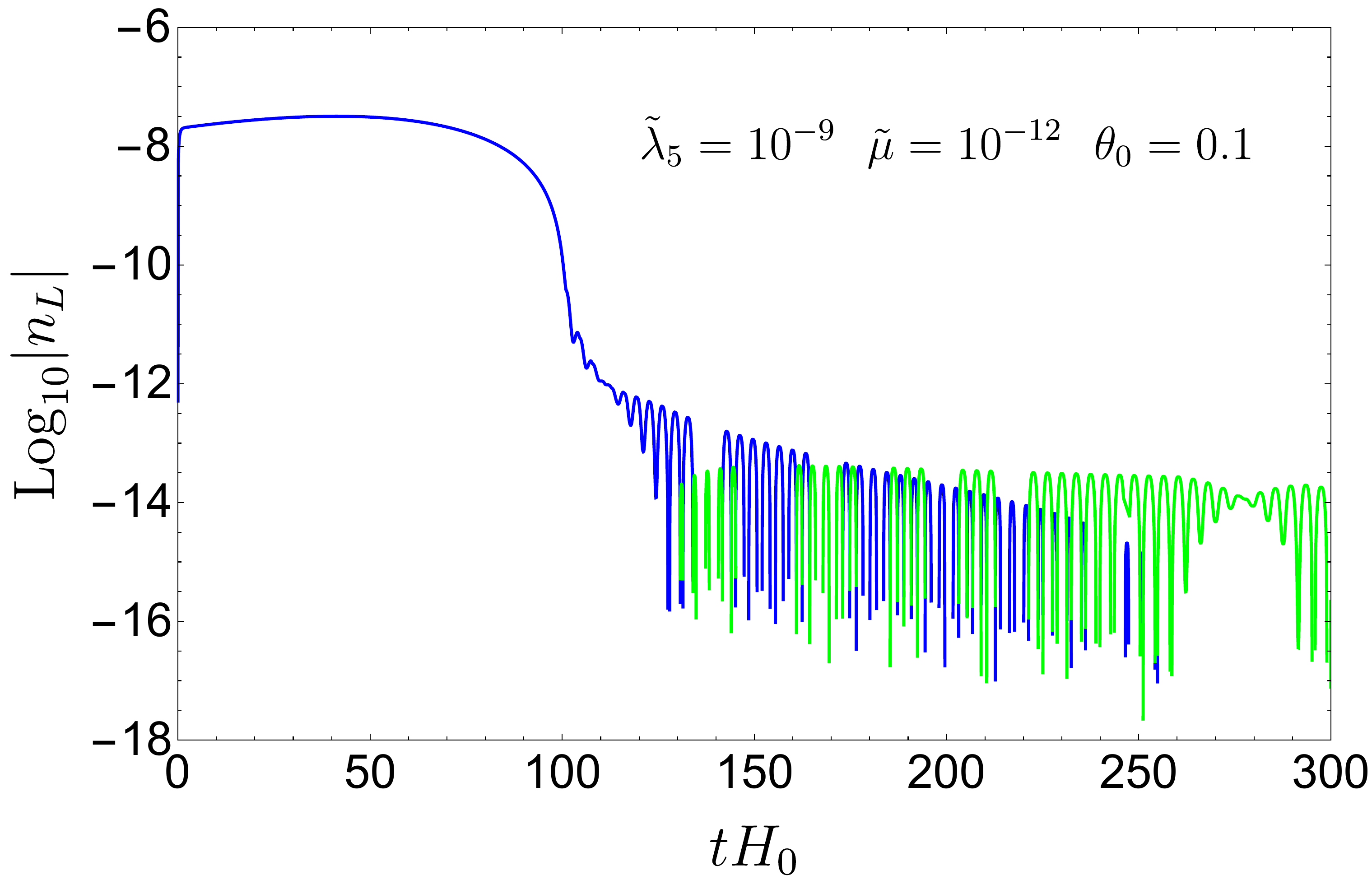}
      	   	\end{subfigure}
      	   	      \begin{subfigure}
      \centering
      	\includegraphics[width=0.491\textwidth]{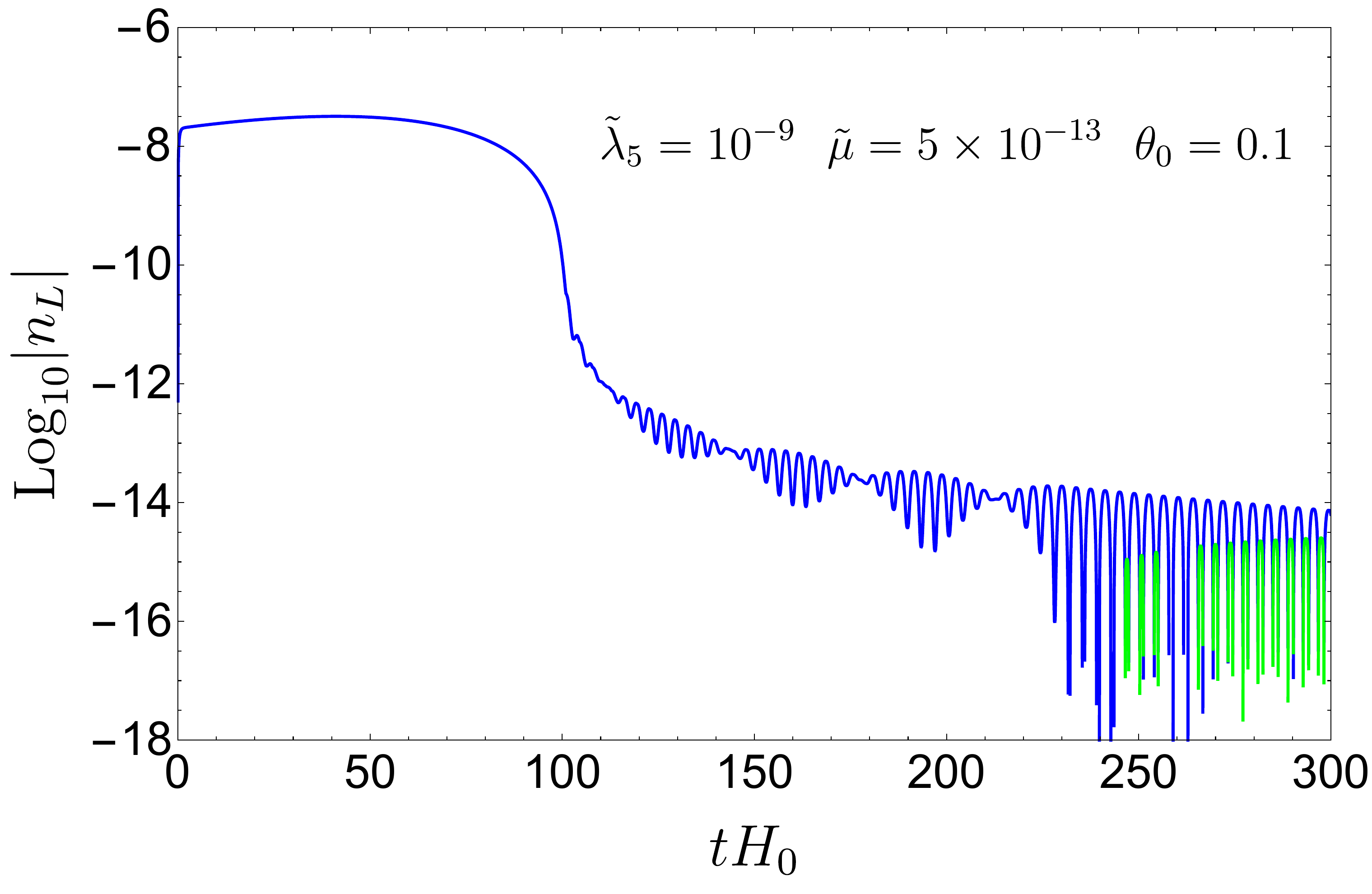}
      	\end{subfigure}
      	\hfill 
      	   \begin{subfigure}
      	  	\centering
      	   	\includegraphics[width=0.491\textwidth]{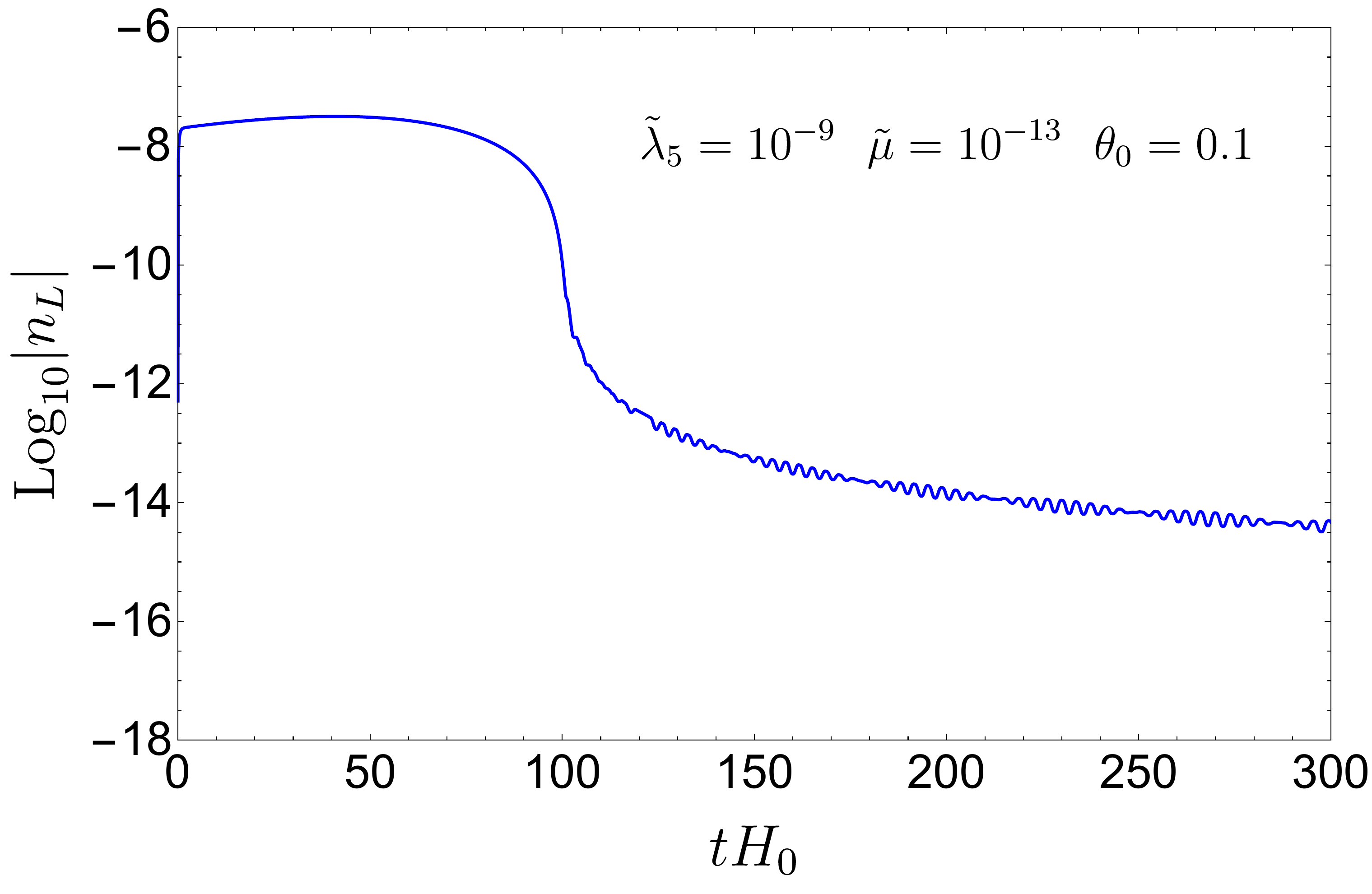}
      	   	\end{subfigure}
      	\caption{  The lepton number density during inflation and and the oscillatory epoch for varying values of the cubic coupling term $\tilde{\mu}$, where we define $H_0= m_S/2$.  The blue and green curves denote positive and negative lepton number densities respectively, and  $ n_L $ and $\tilde{\mu}$ are given in Planck units. The input parameters are fixed to $ \xi= 300$ and $\lambda=4.5 \cdot 10^{-5}$, with initial conditions  $\chi_0=6.0 M_p$, $\dot \chi_0=0$,  and $\dot \theta_0=0 $ chosen.}
      	\label{nL_mu2}
      \end{figure}

We now provide a discussion of the dynamics of the cubic coupling $\tilde \mu$.  Unlike the quartic  and quintic terms, the cubic term becomes increasingly relevant as $\varphi$ decreases. If the cubic term is dominant during the oscillatory phase,  the $\theta$ field will not exhibit directed motion in phase space, and will oscillate instead. This means that the lepton asymmetry starts to oscillate and the predictability of our model is lost. This is depicted in the top-left panel  of Figure \ref{nL_mu2}, where the $U(1)_L$ breaking term is taken to consist only of the $\tilde \mu$ term. 
 
 We must determine the condition that preserves the lepton asymmetry and does not allow this oscillatory behaviour to become significant. During the matter-like epoch, the lepton number density generated from the cubic term during one oscillation time $1/m_S$ can be given approximately by, 
 \begin{equation}
 \Delta n_L\approx \frac{2 Q_L \tilde{\mu} \varphi^3}{m_S}~.
 \label{Dbaryon}
 \end{equation}
 
 To preserve the lepton number density generated during inflation, it is necessary to ensure that $\Delta n_L \lesssim n_L$ throughout the matter-like epoch, prior to the completion of reheating. That is,
 \begin{equation}
 {\tilde \mu \lesssim \frac{m_S {n_L}_{\rm reh}}{4 \varphi_{\rm reh}^3}~.}
 \end{equation}
 
Numerically, we find the following constraint  $\tilde \mu \lesssim 10^{-13} M_p$ when requiring $\Delta n_L \lesssim n_L$, where we have made the  parameter choices $\theta_0=0.1$, $\tilde \lambda_5=10^{-9}, $ and subsequently $ {n_L}_\textrm{end}=1.1\cdot 10^{-11} M_p^{3}$. In Figure \ref{nL_mu2}, we depict the evolution of the lepton asymmetry number density for different values of $\tilde \mu$. It is found that when $\tilde \mu < 10^{-13} M_p $ the lepton asymmetry does indeed become stable, showing consistency with the analytical constraint. For parameter values that lead to the observed baryon asymmetry today, we obtain the following upper bound $\tilde \mu \lesssim 10^{-18} M_p$.


\subsection{The Predicted Baryon Asymmetry}

 Once reheating is completed, the generated non-zero lepton number density $ n_L $ will be present in the form of neutrinos. The lepton number carried by these neutrinos will be redistributed into baryons through the equilibrium electroweak sphalerons, with the ratio defined by $ n_B\simeq -\frac{28}{79} n_L $ \cite{Klinkhamer:1984di,Kuzmin:1985mm,Trodden:1998ym,Sugamoto:1982cn}. 
 To calculate the baryon asymmetry parameter generated in our scenario, we must determine the reheating temperature.
 
 The process of reheating in standard Higgs inflation was first analysed in Ref. \cite{Garcia-Bellido:2008ycs, Bezrukov:2008ut}, where it was found that the parametric resonance production of $W/Z$ bosons plays a  significant role in the preheating process. It has since been found that the preheating process is more violent than was previously expected \cite{Ema:2016dny, DeCross:2015uza, DeCross:2016cbs, DeCross:2016fdz}, with unitarity being violated for SM Higgs non-minimal couplings of $\xi>350$. However, recently it has been found that for such large $\xi$ values, it is possible for the model to be UV complete in Higgs-$R^2$ scenarios of inflation \cite{Giudice:2010ka,Gorbunov:2018llf, Ema:2017rqn, Ema:2019fdd, Ema:2020zvg} for which the preheating process must be recalculated  \cite{He:2018mgb,He:2020ivk,He:2020qcb}. In our scenario, we choose $\xi=300$ to avoid any possible issues with unitarity. A recent analysis of the preheating process in SM Higgs inflation  \cite{Sfakianakis:2018lzf} has shown that for models with $\xi > 100$, the reheating happens at approximately $\sim 3$ e-folds after the end of inflation, so we adopt $\Delta N= N_\textrm{reh}- N_\textrm{end} = 3$ for simplicity. The details of reheating may be different in our scenario due to the mixing of the SM and triplet Higgs', and thus, it is necessary to undertake a comprehensive analysis, which will be left to future work.
 
For the typical parameters chosen, the reheating process occurs at $t_\textrm{reh} =223/H_0$, which has the corresponding Hubble parameter $H_\textrm{reh}=0.0047 H_0$. From the relation $ H_\textrm{reh}^2\simeq \frac{\pi^2}{90} g_* \frac{T_\textrm{reh}^4}{M_p^2} $, we then obtain the reheating temperature 
$T_\textrm{reh} \approx 2.2\cdot10^{14}$ GeV. Considering the entropy density just after reheating defined by  $s=\frac{2 \pi^2}{45} g_* T_\textrm{reh}^3 $ \cite{Husdal:2016haj}, we then determine that the baryon asymmetry parameter generated in our model is given by,
\begin{equation}
\eta_B = \frac{{n_B}}{s}  \bigg|_{\rm reh} = \eta_B^\textrm{obs} \left(\frac{{|n_L}_\textrm{end}|/M_p^{3}}{1.3\cdot 10^{-16} }  \right) \left( \frac{g_*}{112.75} \right)^{-\frac{1}{4}}~,
\label{etabaryon1}
\end{equation}
where $\eta_B^\textrm{obs} \simeq 8.5 \cdot 10^{-11}$ is the observed baryon asymmetry parameter \cite{Aghanim:2018eyx}. This relation shows that at the end of inflation, it is necessary to have generated a lepton number asymmetry density of $1.3\cdot 10^{-16} M_p^{3}$ to produce the observed  baryon asymmetry, which corresponds to the example parameter sets:  $\tilde{\lambda}_5 = 7\cdot 10^{-15}$ for $\theta_0=0.1$ and $\tilde{\lambda}_5 = 10^{-10}$ for $\theta_0=6.5 \cdot 10^{-6}$ from numerical calculations. These example parameters illustrate the wide range of parameters over which successful Leptogenesis can occur. Note that in both of these example cases, the typical parameters escape the isocurvature constraints \cite{Byrnes:2006fr, Gordon:2000hv, Kaiser:2012ak} placed by CMB observations \cite{Planck:2018jri}; see the next section for the detailed calculations of the expected isocurvature perturbations in our framework.

As detailed in the previous subsection, if the $\tilde \mu$ coupling is too large,  the lepton asymmetry will begin to rapidly oscillate during the matter-like epoch and the predictability of our model breaks down. On the other hand, a smaller $\tilde \mu$ term helps to avoid the washout of the lepton asymmetry after reheating. In the following, we require $|\tilde \mu| \lesssim 10^{-18} M_p$ for the initial $\theta_0=0.1$ to ensure the generated baryon asymmetry is preserved and consistent with observation.

In our analysis, we choose the parameter inputs $\lambda_H =0.1$, and  $\xi_H =\xi_\Delta= 300$ based on the arguments laid out above. For the other parameters, we  set $\lambda_{\Delta}= 4.5\cdot 10^{-5}$ to accommodate the current CMB measurements, while there exists two options for $\lambda_{H \Delta}$ that ensure the required inflationary trajectory is preserved. In the case of  $\lambda_{H \Delta} > 0$, we require $2\lambda_\Delta \xi_H-\lambda_{H\Delta}\xi_\Delta>0$ and $2\lambda_H \xi_\Delta-\lambda_{H\Delta} \xi_H>0$ so that we have a mixed state of $h$ and $\Delta^0$ as the inflaton. The typical value we can consider is $\lambda_{H \Delta}=10^{-5}$, which gives the mixing angle as $\alpha \simeq 0.02$. In the case of $\lambda_{H \Delta} < 0$, we require that $|\lambda_{H \Delta}| < 2 \sqrt{\lambda_H \lambda_{\Delta}}$ to avoid the potential becoming unbounded from below. We can then choose  $\lambda_{\Delta}= 4.5\cdot 10^{-5}$, $\lambda_{H\Delta}=-0.001$, which gives $\alpha \simeq 0.07$. 

We stress here that the above parameter choices are just a simplified version of our benchmark model considered in previous subsections. For the parameters in a real model, we need to also consider the contributions of radiative corrections, as well as the satisfaction of the vacuum stability conditions. As we have mentioned before, the selection of  $\lambda_{H\Delta}\sim \lambda_\Delta \sim \mathcal{O}(0.01-0.1)$  may be a more realistic coupling choice to avoid any issues that could arise from radiative corrections. 

Numerically, an example set of parameters that successfully generates the observed baryon asymmetry is $\lambda_5 \sim \lambda_5^{\prime}= 8.8\cdot 10^{-11} ~(7.9\cdot 10^{-12}) $ with $\theta_0\sim 0.1$ in combination with the typical parameter choices given above, for the  $\lambda_{H\Delta}>0$ ($\lambda_{H\Delta}<0$) cases respectively. In addition, we obtain an upper limit on the cubic term coupling of $|\mu|\lesssim 15 ~(1.4)$ TeV.  Note that, all these parameters are defined at a renormalization scale near $M_p$. Given that the $U(1)_L$ breaking terms exhibit small couplings, it is natural to consider that they originate from a spurion field that carries a $U(1)_L$ charge of $+2$. The $U(1)_L$ breaking terms are subsequently generated by requiring that the spurion field obtains a VEV of the order of $\mathcal{O}(10^{4})$ TeV. 


\subsection{Lepton Number Washout Effects}  

Since we expect a large reheating temperature in this scenario, the triplet Higgs will rapidly thermalize at the beginning of the radiation epoch, and so we must consider possible washout processes. Firstly, we require that the process $LL\leftrightarrow  HH$ is never in thermal equilibrium,
\begin{equation}
\Gamma|_{T=m_\Delta}=n\langle\sigma v\rangle\approx y^2 \mu^2/m_\Delta< H|_{T=m_\Delta}~,
\end{equation}
where $ H|_{T=m_\Delta}=\sqrt{\frac{\pi^2 g_*}{90}}\frac{m_\Delta^2}{M_p}$. 

Remember that the triplet Higgs  generates the neutrino masses through the following definition $m_\nu\simeq y\frac{|\mu| v^2}{2m_\Delta^2}$, and that there should be at least one neutrino mass at least of the order of the $m_\nu\sim 0.05$ eV. Combining this with the above relation, we obtain a limit on the triplet Higgs mass of $m_\Delta <10^{12}$ GeV for $m_\nu=0.05$ eV. 

The other potentially dangerous processes that we must consider are $LL\leftrightarrow\Delta$ and $ HH\leftrightarrow\Delta $. They must not co-exist, otherwise, the lepton number asymmetry will be rapidly washed out. Following Ref. \cite{Harvey:1990qw}, one can easily find that the only solution for the corresponding chemical potentials, if both of these processes are in equilibrium, is $\mu_L= \mu_{u_L}= \mu_H=0$. However, to maintain the lepton asymmetry generated during inflation, the process $LL\leftrightarrow\Delta$ must be efficient while  $HH\leftrightarrow\Delta $ is out of equilibrium. This condition can be easily realised for a moderate  $m_\Delta$ and a sizable $y$, for example, $m_\Delta < 10^8$ GeV and $y> 10^{-5}$. 
The requirement to ensure that these two processes do not coexist is given by,
\begin{equation}
\Gamma_{ID}(HH\leftrightarrow \Delta)|_{T=m_\Delta}<H|_{T=m_\Delta}~.
\label{con1}
\end{equation}
where 
\begin{equation}
\Gamma_{ID}(HH\leftrightarrow \Delta)|_{T=m_\Delta} \approx \Gamma_{D}(\Delta \rightarrow HH) \simeq \frac{\mu^2}{32\pi m_\Delta}~.
\end{equation}

Using that $v_\Delta \simeq - \frac{\mu v_{\textrm{EW}}^2}{ 2 m^2_\Delta}$ with Eq.~(\ref{con1}), the necessary condition is found to be,
\begin{equation}
v_\Delta \lesssim 10^{-5}~{\rm GeV} \left( \frac{m_\Delta}{1~{\rm TeV}}\right)^{-1/2} ~,
\label{vev1}
\end{equation}
hence, for $m_\Delta \gtrsim 1$ TeV, we generally require that $v_\Delta \lesssim 10$ keV to prevent the washout of the lepton asymmetry. This is an important feature of our model with significant phenomenological implications, as will be discussed in the following section.


\section{Phenomenological and Cosmological Implications}
\label{Phenom}

The Type II Seesaw Mechanism alone has rich phenomenological implications both terrestrially and cosmologically.  The inclusion of the associated triplet Higgs in the inflationary setup and its responsibility for the production of the matter-antimatter asymmetry places unique constraints on the allowed parameter space of the model, and allows for additional avenues for experimental verification. Here we will discuss some of these possibilities, and provide a summary of the parameter space for successful Baryogenesis that is consistent with each of these constraints. We shall also consider the detection prospects from future experimental results.

It should be noted, that although not discussed below, there are potential Gravitational Wave signals that may be associated with our model, in addition to the predicted tensor-to-scalar ratio. Namely, from the dynamics of the preheating process \cite{Figueroa:2017vfa,Caprini:2018mtu}, and from the inducement of a first-order phase transition through the new interactions introduced into the scalar potential \cite{Zhou:2022mlz}.


\subsection{Current and Future Collider Constraints}

Searches for the charged components of the triplet Higgs at terrestrial colliders provide important constraints on its allowed mass. Depending upon the VEV of the triplet Higgs and its Yukawa couplings, they can decay dominantly into  gauge bosons or leptons \cite{Melfo:2011nx}. Generally, for $v_\Delta \lesssim 0.1 $ MeV, the triplet Higgs will dominantly decay into leptons, while for  $v_\Delta \gtrsim 1 $ MeV, it will dominantly decay into gauge bosons. Cascade decays are also possible if the mass splitting between the triplet Higgs components is sizable, namely $\gtrsim$ 10 GeV. However, such large mass splittings require a large $\lambda_4$ coupling. To keep perturbativity and vacuum stability up to $M_p$ scale, the $\lambda_4$ coupling cannot be much larger than $\mathcal{O}(1)$, and subsequently it has been shown that the mass splitting between the different states of the triplet Higgs is generally less than 10 GeV for $m_\Delta > 1$ TeV \cite{Bonilla:2015eha}.   

In our scenario, $v_\Delta$ can only be accommodated within the range  0.05 eV $\lesssim v_\Delta \lesssim $ 10 keV, meaning that the triplet Higgs will dominantly decay into leptons. If  we observed the  triplet Higgs in such a channel, it would provide a smoking gun for our model. In fact, the LHC has already performed searches for the triplet Higgs via decays to the leptonic channel from which a lower limit of $\sim 800$ GeV has been determined for the doubly-charged Higgs \cite{ATLAS:2017xqs}. The future 100 TeV collider could explore larger mass regions, and preliminary studies~\cite{Du:2018eaw} show that a doubly-charged Higgs of mass of up to 4 TeV will be probed at the 5$\sigma$ level.


\subsection{Lepton Flavor Violation  Experimental Tests}

The doubly-charged component of the triplet Higgs is responsible for lepton flavor violating processes such as $\mu \rightarrow e \gamma$ and $\mu \rightarrow e e e$.  In particular, the channel that can be probed to the highest precision is provided by the decay $\mu^+ \to e^+ e^- e^+$ which has the following branching ratio \cite{Dinh:2012bp},
\begin{equation}
{\cal B} (\mu^+ \to e^+ e^- e^+) = \frac{|y_{\mu e} y^\dagger_{ee}|^2}{16 G_F^2 m_{\Delta^{++}}^4}\,~,
\end{equation}  
which can then be compared with the current best upper bound \cite{SINDRUM:1987nra},
\begin{equation}
{\cal B} (\mu^+ \to e^+ e^- e^+) \leqslant 1.0 \cdot 10^{-12}\,~.
\end{equation}

The lepton flavor violating decay process $\mu \to e \gamma$ is generated at one-loop level with the help of the doubly and singly-charged scalar components of the triplet Higgs. As a result, the partial width for this process is given by \cite{Dinh:2012bp},
\begin{equation}
{\cal B} (\mu \to e \gamma) \simeq \frac{\alpha}{3072\pi} \frac{\left|(y^\dagger y)_{e\mu}\right|^2}{G_F^2} \left(\frac{1}{m_{\Delta^{+}}^2} + \frac{8}{m_{\Delta^{++}}^2}\right)^2\,~,
\label{meg}
\end{equation} 

Nowadays, the most stringent constraint on $\mu \to e\gamma$ is given by the MEG collaboration which has determined the following upper bound on this process \cite{MEG:2016leq},
\begin{equation}
{\cal B} (\mu \to e\gamma) < 4.2 \cdot 10^{-13}\,~.
\end{equation}

To translate the full implications of these constraints to the parameter space of our model, it is necessary to perform a thorough parameter scan which includes the current neutrino data. This is beyond the scope of this paper and here we just assume all the couplings are of the same order for simplicity. The current experimental limits on these lepton flavor violating decay processes already provide constraints on the triplet Higgs properties, see Figure \ref{washout1}, with future experiments such as  Mu3e~\cite{Perrevoort:2018ttp} to improve upon the $\mu \rightarrow e e e$ limits by two orders of magnitude. Thus, the allowed parameter space of our model will be tested through these processes in the near future.


\subsection{Neutrino Experiment Tests}

The Type II Seesaw Mechanism can be probed through precision tests of rare processes in the neutrino sector. These are associated with the new couplings introduced between the triplet Higgs and the neutrinos, as well as through the generation of Majorana mass terms for the neutrinos.

Neutrinoless double beta decay is a predicted experimental feature of models containing Majorana mass terms for the neutrinos, which will be probed in the near future. Theories that provide unique predictions may be differentiated or ruled out by increased experimental precision in the measurement of this process. In this model, the observed neutrino masses are of the Majorana type. This is in contrast to  models which include right-handed neutrinos, where the observed neutrinos can have both Dirac and Majorana type mass terms. Thus, our model can be probed in near future neutrinoless double beta decay experiments. 
 
Additionally, the baryon asymmetry generated in this model is independent of the leptonic $\mathcal{CP}$ phase, with $\mathcal{CP}$ spontaneously broken in the leptonic sector at early times in the universe.  There are currently conflicting measurements of the leptonic $\mathcal{CP}$ phase coming from the T2K and NOvA experiments, with T2K disfavouring a $\mathcal{CP}$ conserving angle \cite{T2K:2019bcf}, which is inconsistent with the NOvA result \cite{NOvA:2021nfi}. In this context, our model provides an interesting theoretical possibility for Leptogenesis. 


\subsection{Vacuum Stability}

The current experimentally determined values of the SM parameters strongly suggest that the Higgs vacuum is only metastable \cite{EliasMiro:2011aa,Degrassi:2012ry,Lebedev:2012sy,Salvio:2013rja,Branchina:2014usa,Bezrukov:2014ipa}. This can be seen in  the running of the  renormalization group equations (RGE) of the Higgs quartic coupling $\lambda_h$. It should be noted that the running of $\lambda_h$ is sensitive to the top pole mass, which has yet to be determined with high precision. There have been different models proposed to avoid the metastability problem. One interesting possibility is the introduction of a quintessence coupling to the SM Higgs, which ensures that in the early universe the Higgs self-coupling is kept positive \cite{Han:2018yrk}.

In the context of our model, it is important to consider the change in the renormalization group running for the Higgs self-coupling induced by the introduction of the triplet Higgs, and whether the vacuum can be stabilized through its couplings. Investigations into the implications of the Type II Seesaw Mechanism for vacuum stability have been explored extensively \cite{Arhrib:2011uy,Chun:2012jw,BhupalDev:2013xol,Bonilla:2015eha,Haba:2016zbu,Mandal:2022zmy,Moultaka:2020dmb}. This has resulted in the determination of the following conditions, which ensure vacuum stability up to the Planck scale,
\begin{eqnarray}
&& \mathcal{B}_0=\biggl\{ \lambda_H>0; ~~~~ \lambda_2 +\lambda_3 >0; ~~~~ \lambda_2+\frac{\lambda_3}{2}>0  \biggr\}~,  \\ 
&& \mathcal{B}_1=\biggl\{ \lambda_1 +2 \sqrt{\lambda_H(\lambda_2+\lambda_3)}>0; ~~~~ \lambda_1 +\lambda_4+ 2 \sqrt{\lambda_H(\lambda_2+\lambda_3)}>0; \nonumber \\
&&~~~~~~~~~ 2\sqrt{\lambda_H} \lambda_3  \leq \sqrt{(\lambda_2+\lambda_3)\lambda_4^2}   \biggr\}~,  \\
&& \mathcal{B}_2= \biggl\{ 2\sqrt{\lambda_H} \lambda_3 \geq \sqrt{(\lambda_2+\lambda_3)\lambda_4^2}; \nonumber \\
&&~~~~~~~~ \lambda_1+\frac{\lambda_4}{2} +2\sqrt{\lambda_H(\lambda_2+\frac{\lambda_3}{2})(1-\frac{\lambda^2_4}{8\lambda_H \lambda_3})} >0 \biggr\}~,
\end{eqnarray}
where it is required that $\mathcal{B}_0 \land \{\mathcal{B}_1 \lor \mathcal{B}_2 \}$ be satisfied when imposed at any renormalization scale up to $M_p$ to avoid vacuum instability below the Planck scale \cite{Bonilla:2015eha,Moultaka:2020dmb}. It is important to note that, the widely used older conditions found in Ref. \cite{Arhrib:2011uy,Chun:2012jw,BhupalDev:2013xol,Haba:2016zbu,Mandal:2022zmy} are sufficient but not necessary. Note that, the threshold corrections from the triplet Higgs can be ignored due to the smallness of the $\mu$ term.  In summary, it has been found that there are vast regions of parameter space where the stability condition can be satisfied \cite{Bonilla:2015eha, Moultaka:2020dmb}.


\subsection{Isocurvature Perturbations}

In our scenario, we have assumed that the inflationary period is induced by two scalar fields. Generally, such inflationary setups can generate non-trivial non-Gaussian features ~\cite{Kaiser:2012ak}. 
Additionally, we must consider whether the isocurvature perturbations produced by the inflationary dynamics of $\theta$ are within the current observational limits. In the calculation presented below, we follow the works of Ref. \cite{Cline:2019fxx} and \cite{Kawasaki:2020xyf}, and use the formalism employed in Ref. \cite{Gordon:2000hv, Kaiser:2012ak}.

Besides the $\chi$ field, in our model, the only other relevant dynamics during inflation are from the $\theta$ field which is responsible for the generation of the lepton number asymmetry. We calculate the isocurvature perturbations by considering the dynamics of our two-field inflation framework $(\chi,\theta)$. A general Einstein frame action can be written as follows,
\begin{equation}
S_{\rm E} = \int d^4 x \sqrt{-g} \left[\frac{M_p^2}{2}R- \frac{1}{2} h_{IJ} g^{\mu\nu} \partial_\mu \phi^I \partial_\nu \phi^J- V(\phi) \right]  ~,
\end{equation}
where $h_{IJ}$ is the field space metric.
We make the following definition $\phi^I(x^\mu)= \varphi^I(t)+\delta \phi^I(x^\mu)$, which gives the following equations of motion for the background fields $\varphi^I(t)$,
\begin{equation}
\mathcal{D}_t \dot \varphi^I+3 H \dot \varphi^I + h^{IJ}V_{,K} =0  ~,
\end{equation}
where $\mathcal{D}_t$ is the covariant directional derivative defined by
$\mathcal{D}_t A^I\equiv \dot \varphi^I \mathcal{D}_J A^I = \dot A^I + \Gamma^I_{JK} A^J\dot \varphi^K $.
The gauge invariant scalar perturbations, known as the Mukhanov-Sasaki variables, are defined as,
\begin{equation}
Q^I=\delta \phi^I +\frac{\dot\varphi^I}{H} \psi~.
\end{equation}

Thus, we obtain,
\begin{equation}
\mathcal{D}_t^2 Q^I + 3 H \mathcal{D}_t Q^I + \left[ \frac{k^2}{a^2} \delta^I_J + \mathcal{M}^I_J -\frac{1}{M_p^2 a^3} \mathcal{D}_t \left(  \frac{a^3}{H}  \dot \varphi^I \dot \varphi_J \right) \right] Q^J =0  ~,
\end{equation}
where
\begin{equation}
\mathcal{M}^I_J \equiv h^{I K} (\mathcal{D}_J \mathcal{D}_K V) - \mathcal R^I_{LMJ} \dot \varphi^L \dot \varphi^M  ~.
\end{equation}

The adiabatic field $\sigma$ and its direction  $\hat \sigma^I$ can be calculated using,
\begin{equation}
\dot\sigma^2 = h_{IJ} \dot \varphi^I \dot \varphi^J  ~, ~~~\textrm{and}~~~
\hat\sigma^I\equiv \frac{\dot \varphi^I}{\dot \sigma}~.
\end{equation}

Then we have,
\begin{eqnarray}
&&\ddot \sigma + 3 H \dot \sigma + V_{,\sigma} =0 ~,\\
&& H^2 =\frac{1}{3 M_p^2} \left[  \frac{1}{2} \dot \sigma^2 + V \right]  ~, \\
&& \dot H = -\frac{1}{2 M_p^2} (\dot \sigma)^2  ~,
\end{eqnarray}
where $V_{,\sigma} \equiv \hat \sigma^I V_{,I}$.
We then define the entropy direction $\hat s^I$ as,
\begin{equation}
 \hat s^I\equiv \frac{\omega^I}{\omega} ~, ~~~\textrm{and}~~~
 \omega^I\equiv \mathcal{D}_t \hat \sigma^I  ~, 
\end{equation}
where $\omega=\sqrt{h_{I J}\omega^I\omega^J}$. 

The slow-roll parameters are then given by,
\begin{eqnarray}
\epsilon & \equiv& -\frac{\dot H}{H^2} = \frac{3 \dot \sigma^2}{\dot \sigma^2+2V} ~, \\
\eta_{ss} &\equiv& M_p^2 \frac{\hat s_I \hat s^J \mathcal{M}^I_J}{V} ~, \\
\eta_{\sigma\sigma} &\equiv& M_p^2 \frac{\hat \sigma_I \hat \sigma^J \mathcal{M}^I_J}{V}  ~. 
\end{eqnarray}

The gauge invariant adiabatic and isocurvature perturbations are parameterized by,
\begin{equation}
\mathcal{R}_c = (H/\dot \sigma) \hat \sigma_I Q^I ~, ~~~\textrm{and}~~~
\mathcal{S} = (H/\dot \sigma)\hat s_I Q^I ~.
\end{equation}

The evolution of $\mathcal{R}_c$ and $\mathcal{S}$ after horizon crossing is given by,
\begin{equation}
\dot{\mathcal{R}}_c = \alpha H \mathcal S + \mathcal O (\frac{k^2}{a^2 H^2})  ~, ~~~\textrm{and}~~~
\dot{\mathcal{S}} = \beta H\mathcal{S} +\mathcal{O} (\frac{k^2}{a^2 H^2})  ~, 
\end{equation}
where $\alpha =\frac{2\omega(t)}{H(t)}$ and $\beta(t) =-2 \epsilon -\eta_{ss}+\eta_{\sigma\sigma} -\frac{4}{3} \frac{\omega^2}{H^2}$.

The transfer functions can be defined as,
\begin{eqnarray}
\left( 
\begin{array}{c}
    \mathcal{R}_c \\
    \mathcal{S} 
\end{array}
\right)
=
\left(
\begin{array}{cc}
    1 & T_{\mathcal RS} \\
    0 & T_{\mathcal SS} 
\end{array}
\right)
\left(
\begin{array}{c}
    \mathcal{R}_c   \\
    \mathcal{S} 
\end{array}
\right)_* ~,
\end{eqnarray}
with
\begin{eqnarray}
T_\mathcal{RS}(t_*, t) &=&\int^t_{t_*}  dt^\prime 2 \omega(t^\prime) T_\mathcal{SS}(t_*, t) ~,\\
T_\mathcal{SS}(t_*, t) &=&\exp\left[ \int^t_{t_*} dt^\prime \beta(t^\prime) H(t^\prime) \right]  ~.
\end{eqnarray}

The correlation of the curvature and isocurvature modes is typically defined by, 
\begin{equation}
\cos \Delta \equiv {T_\mathcal{RS}}/(1+ T^2_\mathcal{RS})^{1/2} ~.
\end{equation}

A limit can be placed on this parameter from the current Planck data, namely $\cos \Delta \lesssim 0.1 $ \cite{Planck:2018jri}, which is the parameter we wish to derive in our scenario.

Now we have all the ingredients necessary to calculate the isocurvature perturbations in our model. Firstly, the scalars in our scenario are,
\begin{equation}
\phi^1 =\chi,~~ \phi^2=\theta, ~~ h_{IJ} =h_{IJ}(\chi) = 
\left(
\begin{array}{cc}
    1 & 0 \\
    0 &  f(\chi) 
\end{array}
\right)  ~,
\end{equation}
where {$f(\chi) \equiv  \frac{\varphi(\chi)^2 \cos^2 \alpha}{\Omega^2(\chi)}$} .

These give the following,
\begin{equation}
\Gamma^1_{ij} = 
\left( 
\begin{array}{ccc}
 0 & 0  \\
 0 & -\frac{1}{2} f^\prime     
\end{array}
\right),~~~
\Gamma^2_{ij} = 
\left(
\begin{array}{ccc}
 0 & \frac{f^\prime}{2 f} \\
 \frac{f^\prime}{2 f}  & 0     
\end{array}
\right) ~.
\end{equation}

      \begin{figure}
      	\centering
      	\includegraphics[width=0.7\textwidth]{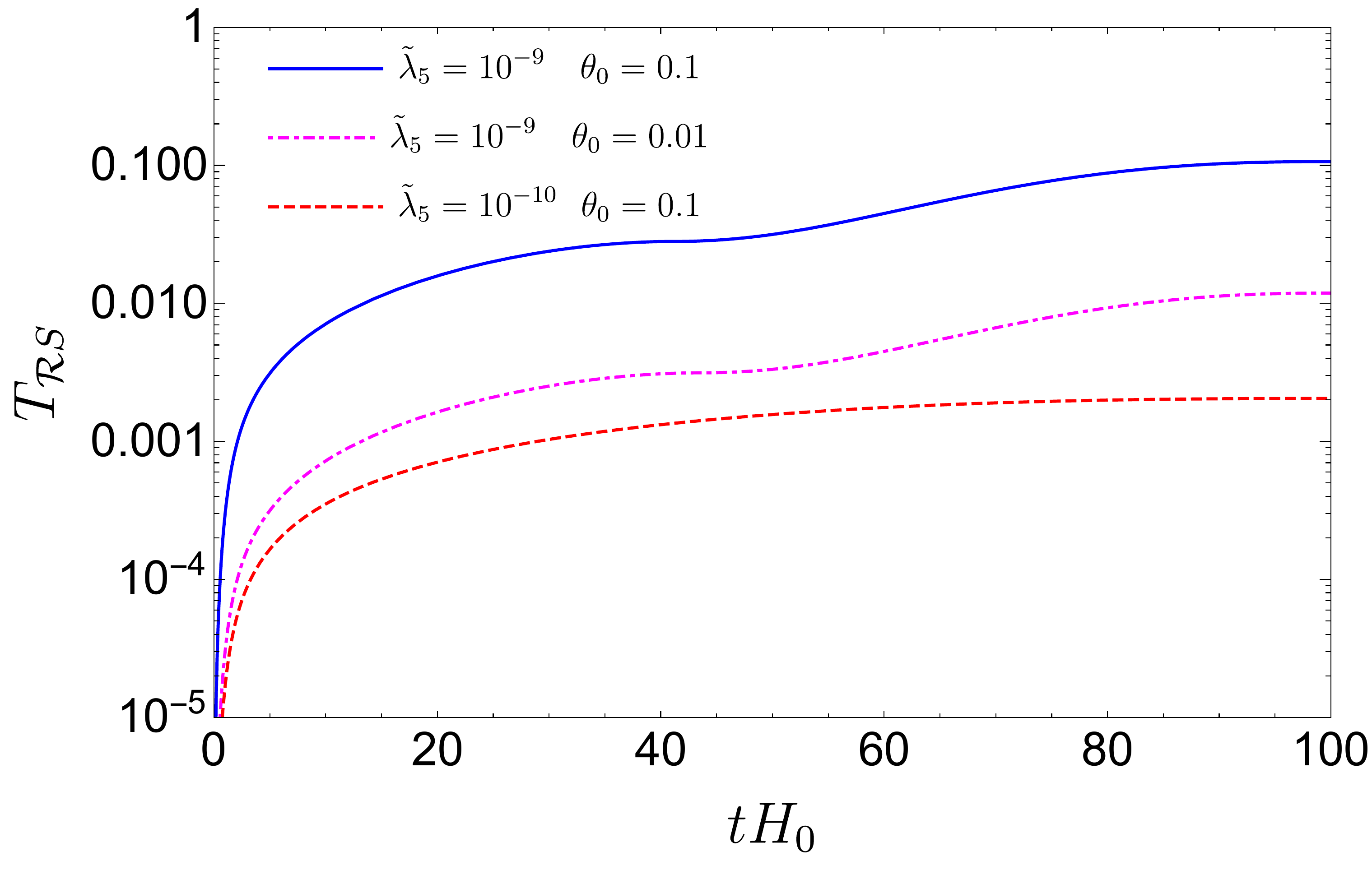}
      	\caption{The evolution of $T_{\mathcal RS}$ during inflation for different parameter values for $\lambda_5$ and $\theta_0$. The other input parameters are fixed to $ \xi= 300$ and $\lambda=4.5 \cdot 10^{-5}$, with initial conditions  $\chi_0=6.0 M_p$, $\dot \chi_0=0$,  and $\dot \theta_0=0 $ chosen. We choose $\tilde\mu$ to be sufficiently small to not affect the dynamics.}
      	\label{TRSa}
      \end{figure}
      
Thus, we find that the only non-vanishing components of the Riemann tensor $R^l_{kji}$ are,
\begin{equation}
R^1_{212} =-R^1_{221} =-R^2_{112} =R^2_{121} = \frac{{f^\prime}^2}{4 f} -\frac{f^{\prime\prime}}{2}   ~,
\end{equation}
which lead to the Ricci curvature tensor $R_{ji}$ and Ricci scalar $R$ as,
\begin{equation}
R_{12} =R_{21} = \frac{{f^\prime}^2- 2 f f^{\prime\prime} }{4f^2} ~, ~~~\textrm{and}~~~
R = \frac{{f^\prime}^2- 2 f f^{\prime\prime} }{2f^2}  ~.
\end{equation}

In Figure~\ref{TRSa}, we depict the numerically calculated evolution of  $T_{\mathcal RS}$ throughout inflation, which must be smaller than $0.1$ to avoid the current observational constraints from the Planck telescope. As exhibited in the figure, the scenario with $\tilde \lambda_5=10^{-9}$ and $\theta_0=0.1$ produces isocurvature perturbations close to the current CMB observational sensitivity. However, this parameter set massively overproduces the observed baryon asymmetry.  It is observed that a reduction of $\tilde \lambda_5$ or $\theta_0$ leads to a smaller $T_{\mathcal RS}$, and thus, successful Leptogenesis in our scenario does not produce sizeable isocurvature perturbations. This is consistent with the expectation that if the parameters are $\tilde \lambda_5=0$ or $\theta_0=0$, there will be no lepton asymmetry generated and the isocurvature mode would disappear.

Thus, a sizable isocurvature signature could be produced only if considerable washout effects are allowed at late times. This possibility may be probed by future observations.


\subsection{Q-ball Formation}
\label{Qball}
In mechanisms for Baryogenesis that involve charged scalar fields, such as the Affleck-Dine mechanism, the formation of Q-balls can occur. If they are produced, it is imperative to investigate their stability and regime of formation to ascertain their phenomenological implications for observations and Baryogenesis. If the  Q-balls are absolutely stable, they can make up part of the dark matter relic density, and in doing so, jeopardise successful Baryogenesis by sequestering the generated asymmetry from the thermal plasma. 

In our model, any Q-balls that are produced will have decay pathways into fermions through the neutrino Yukawa coupling with the triplet Higgs. Thus, as long as the decay time of these processes is such that the Q-balls decay before the EWPT, there should be no phenomenological implications of Q-ball formation during the inflation and oscillation epochs \cite{Coleman:1985ki,Lee:1988ag,Enqvist:2002si,Enqvist:2003gh,Kusenko:1997ad,Rinaldi:2013lsa}. It should be noted that if  Q-Ball formation does occur it can lead to the production of Gravitational Waves which may be probed by future experiments \cite{White:2021hwi}.

We need to first determine if Q-ball formation is possible in our scenario. To do this we must consider  the properties of the effective potential  $U_{\textrm{eff}}(\chi,\omega)=U(\chi)-\frac{1}{2} \omega^2 \chi^2$, and determine the range of $ \omega $ frequencies for which bounce solutions exist. The upper and lower bounds on the frequency, $\omega+$ and $\omega_-$, are given by,
\begin{equation}
\omega_+^2=U''(0)~, \textrm{~~~and~~~} \omega_-^2=\left. \frac{2U(\chi)}{\chi^2}\right|_{\textrm{min}}~,
\end{equation}
where $ U(\chi) $ is given in Eq. (\ref{Uchi_sm1}). These relations lead to the following requirement for the formation of Q-balls,
\begin{equation}
U''(0)>\left. \frac{2U(\chi)}{\chi^2}\right|_{\textrm{min}}~,
\end{equation}
where we have taken $m_\Delta \simeq \sqrt{U^{\prime \prime}(0)}$. 

    \begin{figure*}
	\centering
	\includegraphics[width=0.75\textwidth]{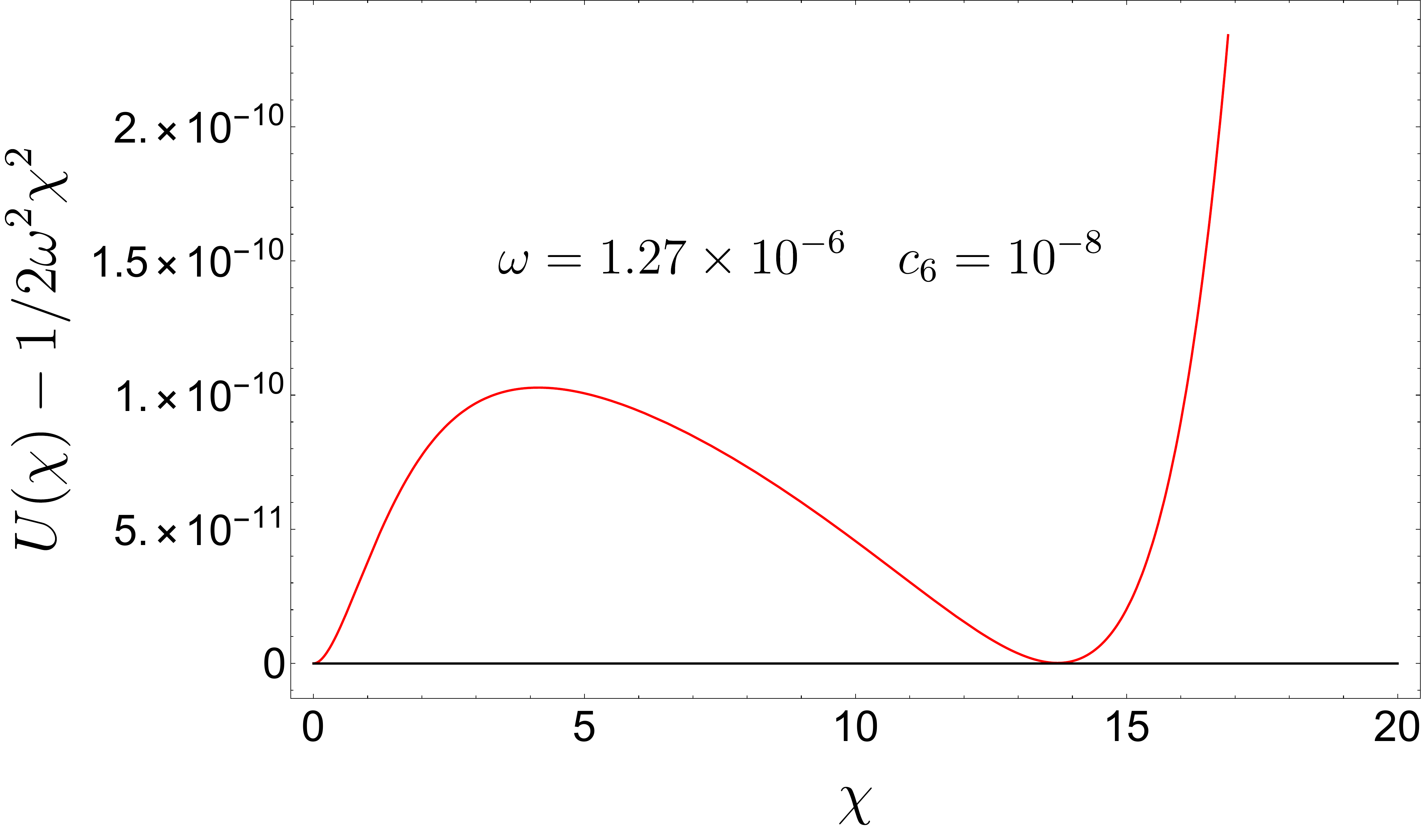}
	\caption{An example of the effective potential for bounce solutions in our framework, with all of the parameters given in Planck units.}
	\label{bounce1}
\end{figure*}

A non-trivial $\omega_-$ is obtained when the potential just obtains two degenerate minima, as depicted in Figure \ref{bounce1}, with one at the origin and the other at a large $\chi$ field value. Given that the potential becomes flat for $\chi\gg M_p$, all the values of $\omega_->0$ will allow for Q-ball formation at sufficiently large $\chi$. However, it is expected that higher-dimensional operators of $\phi$ should exist in the Jordan frame potential. Translating these terms into the canonically normalised Einstein frame field $\chi$, they are exponentially enhanced by factors of $e^{\frac{n}{\sqrt{6}}\frac{\chi}{M_p}}$ depending on the dimensionality of the term, where $n+4$ is the dimension of the operator. A dimension 6 operator of the form ${c_6}{\frac{\phi^6}{M_p^2}}$, increases faster with $\chi$ than the $\omega_-$ dependent term, with the lower bound on $\omega$ then dependent upon the choice of the coupling ${c_6}$. It should be noted that $c_6$ must also be sufficiently small to not destroy the inflationary dynamics of our framework.

An example parameter choice is depicted in Figure \ref{bounce1}, for which $\omega_{-} \simeq 1.27 \cdot 10^{-6} M_{p} $ is the lower bound for a dimension 6 coupling, $c_6=10^{-8}$.  This value of $\omega_{-}$ is much larger than the range of masses $m_\Delta$ that we consider in our scenario. An approximate requirement for Q-Ball formation can then be derived that relates the $c_6$ coupling and the frequency $\omega_-$,
\begin{equation}
c_6 \simeq 430  \frac{\omega_-}{M_p}  e^{-\frac{m_S}{\omega_-}}~.
\end{equation}

Thus, we can see that significant exponential suppression is required to allow Q-ball formation for small frequencies. If we take $\omega_- \sim m_\Delta\sim 1$ TeV, then  it is necessary for $c_6$ to be incredibly tiny, being  much less than $\sim 2\cdot 10^{-13} e^{-3 \cdot 10^{10}}$. Therefore, we conclude that the formation of stable Q-balls is unlikely to occur in our scenario.


\subsection{Summary of Parameter Space and Experimental Reach}
\label{Param}
 
\begin{figure}
 \centering
 \includegraphics[width=0.6375\columnwidth]{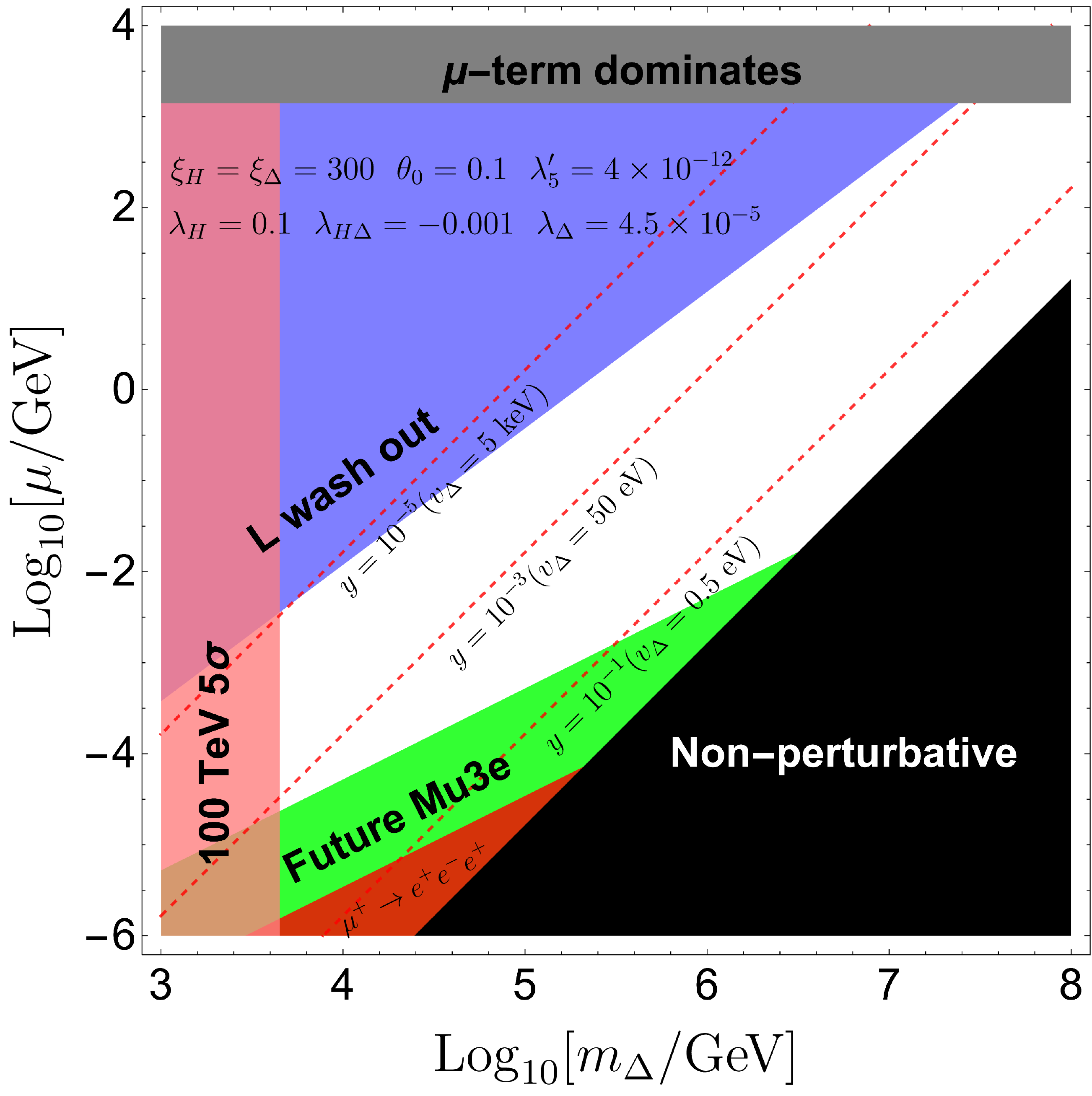}
 \caption{The allowed region of parameter space is depicted (White), avoiding non-perturbative neutrino Yukawa couplings (Black), cubic term domination of the $U(1)_L$ breaking interactions (Grey), and lepton asymmetry washout processes (Blue). The red region indicates the current limits from lepton violating processes \cite{SINDRUM:1987nra}, with green denoting the future Mu3e experimental sensitivity \cite{Perrevoort:2018ttp}. The future 100 TeV collider constraints are shown in the light red region \cite{Du:2018eaw}. We provide dotted red lines to illustrate the corresponding values of the neutrino Yukawa couplings and triplet Higgs VEV within the white allowed region.}
 \label{washout1} 
 \end{figure}

In Figure \ref{washout1}, we show the region of allowed parameter space for which the lepton number density generated during inflation leads to successful Baryogenesis. Requiring perturbative neutrino Yukawa $y$ couplings up to the Planck scale $M_p$ ($y \lesssim 1$) results in the black exclusion region. However, to avoid fine-tuning at the high energy scale, a small Yukawa coupling $y<0.1$ is preferred for the size of quartic coupling we typically consider, namely $\lambda_\Delta\simeq 4.5\cdot 10^{-5}$. To ensure the $\mu$-term does not destroy the generated lepton number asymmetry, as discussed in the previous section, we exclude the grey region. Additionally, significant lepton asymmetry washout effects exist after reheating for parameters within the blue exclusion region. 

We note that the input parameters used in Figure \ref{washout1} are not the only allowed parameter set. For example, we can take $\lambda_{H\Delta}\lesssim \lambda_\Delta\sim 10^{-2}$, and subsequently to obtain the required lepton asymmetry for Eq.~(\ref{nlend1}), one can easily find that the coupling $\lambda_5^\prime$ must be one order of magnitude larger. Simultaneously, in this example, the upper limit on the $\mu$ term, given by the gray region, will be weakened by an order of magnitude. 
 
A limit is obtained from  precision measurements of the triplet Higgs VEV $\langle\Delta^0\rangle$, namely, it must be less than a few GeV. However, this leads to no additional constraints on our model because we require that the VEV must be less than 10 keV to avoid  lepton asymmetry washout effects after reheating. The LHC searches provide lower limits on the masses of the triplet Higgs components, with the current bound on the mass of the doubly charged Higgs being $ \gtrsim 800 $ GeV \cite{ATLAS:2017xqs}, due to which we depict triplet Higgs masses $\geq 1$ TeV in Figure \ref{washout1}. These bounds will be improved at the upgraded high luminosity LHC and at future collider experiments \cite{Du:2018eaw}. 
 
If we assume that each of the neutrino Yukawa couplings are of approximately the same order of magnitude, the lepton flavor violating processes induced by the doubly-charged component of the triplet Higgs already constrain the allowed parameter space  \cite{SINDRUM:1987nra, Han:2021nod}. The current bounds will be improved by two orders of magnitude by the upcoming Mu3e experiment~\cite{Perrevoort:2018ttp}, as indicated in Figure \ref{washout1}.


\section{Conclusions and Future Prospects}
\label{Conc}

The Type II Seesaw Mechanism provides a minimal framework in which to explain the neutrino masses through the introduction of a single triplet Higgs to the SM. It has been believed  that this simple extension cannot provide a satisfactory solution to the unknown origin of the observed baryon asymmetry of the universe, in contrast to the Type I or Type III Seesaw Mechanisms. In this paper, we have demonstrated that the introduction of the  triplet Higgs of the Type II Seesaw Mechanism  can provide a natural way to generate the baryon asymmetry if it also plays a role in setting up the inflationary dynamics. In doing so, a successful and economic Affleck-Dine Baryogenesis scenario is realised without the need for Supersymmetry.  Beyond these successes, the  model exhibits an intriguing and unique combination of phenomenological predictions. We now summarize these possibilities and their future experimental prospects:

\begin{itemize}
	\item  
	The dominant decay processes exhibited by the triplet Higgs is dependent upon the VEV of the triplet and its Yukawa couplings. These decays can be mainly into  gauge bosons or leptons. In our framework,  the $v_\Delta$ is constrained to be within the range of 0.05 eV - 10 keV, with this upper bound required to ensure that the lepton number asymmetry is unaffected by lepton washout effects after reheating. Importantly, within this range of $v_\Delta$ values, the triplet Higgs dominantly decays into leptons. If  we were to observe the  triplet Higgs in the leptonic channel, it would provide a smoking gun for our model. 
	
	\item The doubly-charged component of the triplet Higgs is responsible for lepton flavor violating processes such as $\mu \rightarrow e \gamma$  and $\mu \rightarrow e e e$. The current experimental limits on these processes are already able to place interesting  constraints on the allowed properties of the triplet Higgs in our model, depicted in Figure \ref{washout1}. Planned future experiments, such as  Mu3e, will improve upon the $\mu \rightarrow e e e$ bounds by approximately two orders of magnitude. Thus, one of the most interesting regions of the allowed parameter space of our model will be able to be tested in the near future.
	
	\item The neutrino masses generated in the Type II Seesaw Mechanism are of the Majorana type. This is in contrast to models that include right-handed neutrinos, in which the observed neutrinos exhibit both Dirac and Majorana type mass terms. The presence of the Majorana mass term in our model leads to rare processes such as neutrinoless double beta decay. Thus, the upcoming experiments that will seek to observe neutrinoless double beta decay will be essential tests of our model. 
	
    \item 
	The Leptogenesis scenario we present in this work is independent of the leptonic $\mathcal{CP}$ phase, with the leptonic $\mathcal{CP}$ symmetry instead being spontaneously broken during the early universe. The most up-to-date experimental measurements of the leptonic $\mathcal{CP}$ phase, from the T2K and NOvA experiments, are in conflict. The T2K result disfavours a $\mathcal{CP}$ conserving angle \cite{T2K:2019bcf}, which is inconsistent with the NOvA measurement \cite{NOvA:2021nfi}. In this context, our model provides an interesting theoretical possibility for successful Leptogenesis. 
		
	\item 	
	Gravitational Waves may play an important role in verifying this Leptogenesis mechanism. We predict that the tensor-to-scalar ratio  signal from inflation is $ 0.0033<r<0.0048 $, which may be probed by the near future telescope LiteBIRD  \cite{Hazumi:2019lys}. Also, the preheating process may also provide additional sources of Gravitational Waves with characteristic signatures \cite{Figueroa:2017vfa,Caprini:2018mtu}.
	
	\item The inflationary epoch in our model is set up by the mixture of two scalar fields, the SM and triplet Higgs'. Generally, such inflationary scenarios are able to generate non-trivial non-Gaussian features that can be observed in measurements of the CMB \cite{Kaiser:2012ak}. In addition, it is possible that sizable isocurvature  signatures are produced by our Leptogenesis mechanism  if considerable lepton asymmetry washout effects are allowed to occur after reheating. These possibilities will be probed by future observational efforts. 
\end{itemize}

\acknowledgments
We would like to thank Tsutomu T. Yanagida, Misao Sasaki, Shi Pi and Jiajie Ling for their helpful discussions. C. H. is supported by the Guangzhou Basic and Applied Basic Research Foundation under Grant No. 202102020885, and the Sun Yat-Sen University Science Foundation. NDB was supported by IBS under the project code, IBS-R018-D1.  The work of HM was supported by the Director, Office of Science, Office of High Energy Physics of the U.S. Department of Energy under the Contract No. DE-AC02-05CH11231, by the NSF grant PHY-1915314, by the JSPS Grant-in-Aid for Scientific Research JP20K03942, MEXT Grant-in-Aid for Transformative Research Areas (A) JP20H05850, JP20A203, by WPI, MEXT, Japan, and Hamamatsu Photonics, K.K.

\bibliographystyle{JHEP}
\bibliography{bibly}
\end{document}